\documentclass[12pt]{article}
=24cm
\usepackage{amsmath}
\usepackage{floatflt}
\usepackage{amsfonts}
\usepackage{slashed}
\usepackage{hyper}
\usepackage{cite}
\usepackage{mathrsfs}
\usepackage{bbm}
\usepackage{acronym}
\usepackage{amssymb}
\usepackage{graphicx}%
\numberwithin{equation}{section}
\begin{document}

\noindent {\bf \large  Affine $A^{(1)}_{3}$ $N=2$ Monopole as the D
Module and Affine ADHMN Sheaf }

\bigskip

{\large Hou Bo-Yu } \footnote{ E-mail:  byhou@nwu.edu.cn}$^{,a}$
       and  {\large
Hou Bo-Yuan }$^{b}$

\bigskip

$^{a}$Institute of Modern Physics, Northwest University, Xi'an,
710069, China\newline

$^{b}$Graduate School, Chinese Academy of Science, Beijing 100039,
China\newline

\bigskip

\emph{\large\textbf{Abstract}} \emph{ A Higgs-Yang Mills monopole
scattering spherical symmetrically along light cones is given. The
left incoming anti-self-dual $\alpha$ plane fields are holomorphic,
but the right outgoing SD $\beta$ plane fields are antiholomorphic,
meanwhile the diffeomorphism symmetry is preserved with mutual
inverse affine rapidity parameters $\mu$ and $\mu^{-1}$. The Dirac
wave function scattering in this background also factorized
respectively into the (anti)holomorphic amplitudes. The holomorphic
anomaly is realized by the center term of a quasi Hopf algebra
corresponding to an integrable conform affine massive field. We find
explicit Nahm transformation matrix(Fourier-Mukai transformation)
between the Higgs YM BPS (flat) bundles (D modules) and the
affinized blow up ADHMN twistors (perverse sheafs). Thus establish
the algebra for the 't Hooft-Hecke operators in the Hecke
correspondence of the geometric Langlands Program.}
\bigskip



\bigskip
\bigskip

\section{Introduction}

The 't Hooft BPS monopole always plays important role in gauge field
theory, for example, for the Seiberg Witten monopole condensation
$^{[1]}$ and for the 't Hooft operator $^{[2]}$ . In this paper, we
first give an affine monopole which twist affinize the twistor . The
key points lies in that, when we do the transformation from the
space time fixed frame to the comoving frame, in fact we have
transform to the static $\kappa$ symmetric Killing gauge in Green
Schwarz theory. Meanwhile, we start from the BPS Higgs Yang-Mills
bundle, which is flat over the self dual plane $\alpha$ and the
antiself dual plane $\beta$, to the affinized ADHMN sheaf (i. e.
twistor sheaf).

Then we solve the Dirac equation in this background. Here we
covariantly transform to the same comoving frame. This will manifest
the Nahm transformation to the twistor space. These are realized as
the soliton solution of the conformal affine Toda field $^{[3,4]}$.
Furthermore, this will give the elementary building block for the
moduli space of exactly solvable worldsheet theory (Bena,
Polchinski, Roiban$^{[5]}$, and$^{[6,7]}$). Our work shows the
explicit links between Yang-Mills theory, quantum massive integrable
field and quasi Hopf (Drinfeld$^{[8]}$) twistors for the scaled
elliptic algebra and the scaled W algebra $^{[9-14]}$. So it
generalize the algebraic formulation in the Langlands program by E.
Frenkel et al.$^{[15-17]}$ to the 4 dimensional case.

\section{The affine monopole solution}

\label{sec:cp1} As Kapustin and Witten$^{[2]}$ we consider an affine
family of the N=2 supersymmetric gauge field in the 4d Minkowski
space $\mathbb{M}$ which has been twisted relevant to the geometric
Langlands program. We pick a homotopic homomorphism $\mathfrak{K}$
from the space time symmetry $Spin(4)_{S}$, which is the universal
cover $SU(4)$ of the conformal
$SO(4,2), $ to the $R$ symmetry $Spin(6)_{R}$, $\mathfrak{K}$: $Spin(4)_{S}%
\rightarrow Spin(4)_{R}\subset Spin(6)_{R}$. By the choice of these
Bochner Martinelli kernel homotopy operator $\mathfrak{K},$ we get a
family of N=2 loop supersymmetry $\widetilde{SU}(4)$ with affine
parameter $t\in \mathbb{CP}_{1}.$ To establish the Hecke
correspondence, it should be further central extended to
$A_{3}^{(1)}$. The time reverse and orientation reversal symmetry
described by [2] implies, and the global Riemann Hecke
correspondence requires that the coordinates $x_{\mu}(\mu=0,1,2,3)$
of $\mathbb{M}$ has to be complexifield, embedded into
$\mathbb{C}^{4}$ by analytically continue to the upper and lower
complex planes respectively, such that the connection of the $D$
module becomes flat in the SD planes $\alpha$
and the ASD planes $\beta.$%

\begin{equation}
\lbrack D_{i+},D_{\perp i+}]+{\ast}[D_{i+},D_{\perp i+}]=0,\text{
\ \ \ \ \ \ \ \ \ \ \ \ (}\alpha\text{)\ } \label{1}%
\end{equation}%
\begin{equation}
\lbrack D_{\bar{i}-},D_{\perp\bar{i}-}]-{\ast}[D_{\bar{i}-},D_{\perp\bar{i}%
-}]=0.\text{ \ \ \ \ \ \ \ \ \ \ \ \ \ \ (}\beta\text{)\ \ \ } \label{2}%
\end{equation}

Remark: The flat connection introduced by C.N.Yang $^{[30]}$ is the
$i=1$ case of (\ref{1}), and (\ref{2.5}).

Here in the covariant derivative
\begin{equation}
D_{\mu}=\partial_{\mu}+A_{\mu},
\end{equation}
we have
\begin{equation}
A_{\mu}=A_{\mu}^{\hat{a}}T^{\hat{a}},\mu=0,1,2,3,
\end{equation}
where $T^{\hat{a}}$ is the $A_{3}^{(1)}$ generators, the $\ast$ denotes the
$4$-dimensional Hodge dual, and the tangent vectors $\partial$ in the
\textbf{fixed null frame} for the $\alpha$ planes are%

\begin{align}
\partial_{i_{+}}  &  \equiv\frac{1}{\sqrt{2}}(\frac{\partial}{\partial x^{i}%
}+\frac{\partial}{\partial x^{0}}),\partial_{\perp i_{+}}\equiv\frac{1}%
{\sqrt{2}}(\frac{\partial}{\partial x^{j}}+\mathrm{i}\frac{\partial}{\partial
x^{k}}),\label{2.5}\\
\partial_{i_{-}}  &  \equiv\frac{1}{\sqrt{2}}(\frac{\partial}{\partial x^{i}%
}-\frac{\partial}{\partial x^{0}}),\partial_{\perp i_{-}}\equiv\frac{1}%
{\sqrt{2}}(\frac{\partial}{\partial x^{j}}-\mathrm{i}\frac{\partial}{\partial
x^{k}}),
\end{align}
while the tangent vectors $\overline{\partial}$ for the $\beta$ planes are%
\begin{align}
\overline{\partial}_{i_{-}}  &  \equiv\frac{1}{\sqrt{2}}(\frac{\partial
}{\partial\overline{x^{i}}}-\frac{\partial}{\partial\overline{x^{0}}%
}),\ \overline{\partial}_{\perp i_{-}}\equiv\frac{1}{\sqrt{2}}(\frac{\partial
}{\partial\overline{x^{j}}}-\mathrm{i}\frac{\partial}{\partial\overline{x^{k}%
}}),\\
\overline{\partial}_{i_{+}}  &  \equiv\frac{1}{\sqrt{2}}(\frac{\partial
}{\partial\overline{x^{i}}}+\frac{\partial}{\partial\overline{x^{0}}%
}),\overline{\partial}_{\perp i_{+}}\equiv\frac{1}{\sqrt{2}}(\frac{\partial
}{\partial\overline{x^{j}}}+\mathrm{i}\frac{\partial}{\partial\overline{x^{k}%
}}).
\end{align}
The $x_{\mu}$ ($\bar{x}_{\mu})$ are the complexified Cartesian coordinate of
$\mathbb{M}$, analytical continued upperwise (lowerwise) respectively. We use
$\perp$ to denote perpendicular in $\alpha$ ($\beta$) plane , $i,j,k$ are the
cyclic permutation of $1,2,3$. The tangential vectors $\partial_{i_{+}}$ and
$\partial_{\perp i_{+}}$ are left null, while $\partial_{i_{-}}$ and
$\partial_{\perp i_{-}}$ are right null. To establish the Hecke
correspondence, we should have left (right) D operators flat on $\alpha$
($\beta$) plane and act on left (right) Hilbert space $\mathcal{H}_{L,R}$
respectively (c.f. the section 3). In this paper we consider the level one
case, i.e. $\mathcal{H}_{L,R}$ are $\left\vert \Lambda_{i}\right\rangle
,\left\langle \bar{\Lambda}_{i}\right\vert $ $(i=0,1,2,3)$ respectively (c.f.
the section 4). Over the $\alpha$ plane (1)$,$ the gauge field $\mathcal{F}%
_{\mu\nu}$ of the left bundle is anti-self dual, while over the $\beta$ plane
(2), $\mathcal{F}_{\bar{\mu}\bar{\nu}}$ is self dual. Here $\mu\nu$ or
$\bar{\mu}\bar{\nu}$ implies that $x_{\mu}$ approach the real slice from upper
or lower half complex planes of $x_{\mu}.$ We will find the monopole solution
separately for the $\alpha$ ($\beta$ ) null planes, such that it is
incoming(outgoing) along the left(right) real null lines $\partial_{i_{+}}($
$\bar{\partial}_{i_{-}})$ spherical symmetrically, i.e. we boost the static
BPS monopole (Appendix A) along the incoming(outgoing) real null lines. The
interaction between incoming and outgoing waves at all the scattering points
yields the central extension (c.f the section 4).

The \emph{level one}, i.e. the grade one in principle gradation of the affine
algrbra, Lorentz covariant spherical symmetric ansatz of the nonzero
components of \textbf{incoming} $A_{\mu}^{\hat{a}}$ in the \textbf{tensor
product} form of spherical \textbf{comoving }spacetime and gauge
\textbf{frames} are (cf. Appendix B)
\begin{align}
K_{T^{+}}  &  =\frac{iF(\mathbbm{r})}{\sqrt{2}\mathbbm{r}}\rho^{-1}%
E^{-1},\label{11}\\
K_{T^{-}}  &  =-\frac{iF(\mathbbm{r})}{\sqrt{2}\mathbbm{r}}\rho^{-1}%
E,\label{5}\\
H_{T^{+}}  &  =\frac{-\partial\gamma/\partial\varphi+\cos\theta}%
{\mathbbm{r}\sin\theta}\rho^{-2},\\
H_{T^{-}}  &  =\frac{\partial\gamma/\partial\varphi+\cos\theta}%
{\mathbbm{r}\sin\theta}\rho^{-2},\\
K_{r-}  &  =A_{r-}=G(\mathbbm{r})\rho^{-2},\label{6}\\
A_{r^{+}}^{c}  &  =\zeta(\mathbbm{r}),\\
\text{ }A_{\mu}^{\mathrm{d}}  &  =0, \label{18}%
\end{align}%
\[
D_{\mu}=\partial_{\mu}+A_{\mu},\qquad A_{\mu}=H_{\mu}+K_{\mu},D_{\mu}%
^{(H)}=\partial_{\mu}+H_{\mu},
\]

\noindent where: the cyclic element
\begin{equation}
E=\left(
\begin{array}
[c]{cccc}%
0 & 1 &  & \\
& 0 & 1 & \\
&  & 0 & 1\\
1 &  &  & 0
\end{array}
\right)  ,
\end{equation}
and the Coxeter $\rho$ lies in the center of $U(4)$
\begin{equation}
\rho=\left(
\begin{array}
[c]{cccc}%
1 &  &  & \\
& i &  & \\
&  & -1 & \\
&  &  & -i
\end{array}
\right)  ,
\end{equation}
i.e. $\left(  \rho^{-1}\right)  _{i,i}=\left(  \rho^{-1}E\right)
_{i,i+1}=\left(  -\omega\right)  ^{i},\left(  \rho^{-1}E^{-1}\right)
_{i+1,i}=\left(  -\omega\right)  ^{i}$, $\omega\equiv e^{{\frac{2\pi
i}{4}}}$, here $\rho$ $E$ ($E\rho=\omega E\rho$) are the generators
of the affine Heisenberg algebra (group), used for the construction
of the vertex operators in the principle realization of the affine
algebra$^{[24]}$. The \textbf{spherical incoming space time null
frames}
\begin{align}
\mathfrak{e}^{T^{+}}  &  =-\frac{1}{\sqrt{2}}\mathbbm{r}e^{-i\gamma}\left(
d\theta+\mathrm{i}\sin\theta d\varphi\right)  ,\\
\mathfrak{e}^{T^{-}}  &  =\frac{1}{\sqrt{2}}\mathbbm{r}e^{i\gamma}\left(
d\theta-\mathrm{i}\sin\theta d\varphi\right)  ,\\
\mathfrak{e}^{r^{\pm}}  &  =\frac{1}{\sqrt{2}}\mu^{\pm1}(dr\pm dt),~~~~r^{2}%
=x_{1}^{2}+x_{2}^{2}+x_{3}^{2}.
\end{align}
here: $\mathbbm{r}=\mu r^{_{+}}+\bar{\mu}^{-1}\bar{r}^{_{-}}$,
$r^{\pm}$ =$\frac{1}{2}(r\pm t)$,
$\gamma(\theta,\varphi)=\mp\varphi$ in north (south) Wu-Yang gauge.
We have decomposed the connection $A_{\mu}$ into the gauge
connection $H_{\mu}$, which lies in the center $U(1)$ of $U(4)$,~
and the covariant constant independent of ($\theta,\varphi$)
components $K_{\mu}$. The connection $H_{\mu}$ turns to be the
$U(1)$ Dirac monopole component for the 't Hooft monopole. Now the
$K_{\mu}$ in eq. (2.9), (2.10), (2.13) are not connections, but are
tensors with the basis elements $A_{\beta,j}$ (given by $\rho,$ $E)$
in lemma$^{[24]}$ $(14.6)$. For the level one representation
$\left\vert \Lambda_{i}\right\rangle $, in the expression of the
principle realization in affine algebra for the vertex operator
$\Gamma,$ only the level one $j=1$ term in the infinite sum of the
following eq.$^{[24]}$(14.6.7)
contributes.%
\begin{equation}
\Gamma^{\beta}=\left\langle \Lambda_{0},A_{\beta},_{0}\right\rangle
\exp\left(  \underset{j=1}{\overset{\infty}{\sum}}\lambda_{\beta j^{\prime}%
}z^{b_{j}}x_{j}\right)  \exp\left(  -\underset{j=1}{\overset{\infty}{\sum}%
}\lambda_{\beta,N+1-j^{\prime}}^{b_{j}}b_{j}^{-1}z^{-b_{j}}\frac{\partial
}{\partial x_{j}}\right)  . \tag{14.6.7}%
\end{equation}
The $j=1$ term usually write as $E_{ij}^{(1)}$, with the grade $^{\prime
\prime}1^{\prime\prime}$ correspond to the exponent $^{\prime\prime}%
1^{\prime\prime}$ of the loop parameter $z$ in (14.6.7).

The explicit expression of the central term $\zeta$ will be given
later. which will contribute the holomorphic anomaly. However it is
irrelevant for the field strength over $\alpha(\beta)$ plane here.
Actually, the incoming and outgoing $K_{\mu}\gamma^{\mu}$ in reduced
cup product form (section 3) turns to be the Affine Toda Lax
connection generate by the grade $1(-1)$ part of the soliton
generating vertex operator (section 4) in the principle realization
$^{[3,4]}$.

Our ansatz for level -1 nonzero component $A_{\bar{\mu}}^{\hat{a}}$ in
\textbf{the outgoing comoving null frames} over the $\beta$ planes~(\ref{2})
are:%
\begin{align}
K_{\bar{T}^{+}}  &  =\frac{iF(\mathbbm{r})}{\sqrt{2}\mathbbm{r}}\rho
E,\label{25}\\
K_{\bar{T}^{-}}  &  =\frac{iF(\mathbbm{r})}{\sqrt{2}\mathbbm{r}}\rho E^{-1},\\
H_{\bar{T}^{+}}  &  =\frac{\partial\bar{\gamma}/\partial\bar{\varphi}-\cos
\bar{\theta}}{\mathbbm{r}\sin\bar{\theta}}\rho^{2},\\
H_{\bar{T}^{-}}  &  =\frac{-\partial\bar{\gamma}/\partial\bar{\varphi}%
-\cos\bar{\theta}}{\mathbbm{r}\sin\theta}\rho^{2},\\
K_{\bar{r}^{+}}  &  =G(\mathbbm{r})\rho^{2},\\
A_{\bar{r}^{-}}^{c}  &  =\zeta(\mathbbm{r}),\\
\text{ }A_{\bar{\mu}}^{d}  &  =0. \label{31}%
\end{align}

As in the case of the static BPS monopole (Appendix (A13) A(14)), \ to obtain
the field strength from the comoving frame ansatz ~(\ref{11})-(\ref{18}) for
the incoming potential become very simple, by using the generalized
Gauss-Codazzi equations(C9), (C10). Now, only the Dirac potential component is
dependent on $(\theta,\varphi)$ to get its $F_{\mu\nu}^{(H)}$ involves
differential calculation, \ all other $K_{\mu}$ terms becomes $(\theta
,\varphi)$ independent tensor matrices, its contribution to the field strength
are just the tensor products and $r$ directional covariant derivatives. The
holomorphic ASD fields strength over $\alpha$ plane are
\begin{align}
F_{T^{+},r^{-}}  &  =\frac{-F(\mathbbm{r})G(\mathbbm{r})}{\mathbbm{r}}\rho
E^{-1},\\
F_{T^{-},r^{-}}  &  =\frac{F(\mathbbm{r})G(\mathbbm{r})}{\mathbbm{r}}\rho E,\\
F_{T^{+},T^{-}}  &  =-\frac{F^{2}(\mathbbm{r})}{\mathbbm{r}^{2}}\rho
^{-2}-\frac{1}{\mathbbm{r}^{2}}\rho^{-2},\\
F_{T^{+},r^{+}}  &  =-\frac{F^{\prime}(\mathbbm{r})}{\mathbbm{r}}\rho
^{-1}E^{-1},\\
F_{T^{-},r^{+}}  &  =-\frac{F^{\prime}(\mathbbm{r})}{\mathbbm{r}}\rho^{-1}E,\\
F_{r^{+},r^{-}}  &  =G^{\prime}(\mathbbm{r})\rho^{-2},
\end{align}
here $G^{\prime}(\mathbbm{r})=\frac{d}{d\mathbbm{r}}G(\mathbbm{r}),$
$F^{\prime}(\mathbbm{r})=\frac{d}{d\mathbbm{r}}F(\mathbbm{r}).$ \

The anti-self-dual equation
becomes%
\begin{align}
F^{\prime}(\mathbbm{r})  &  =G(\mathbbm{r})F(\mathbbm{r}),\\
G^{\prime}(\mathbbm{r})  &  =\frac{F^{2}(\mathbbm{r})-1}{\mathbbm{r}^{2}}.
\end{align}

Remark: the seemingly mismatch factor $\rho^{2}$ is contributed by the
opposite chirality in the $\gamma_{\mu}D_{\mu}^{(H)}$ \ and $\gamma_{\mu
}K^{\mu}$ (c.f. Section 3 ) under homotopy, which include a $\rho^{2}%
\in\mathbb{Z}_{2}$ factor in the Hodge duality in the target space,
moduli space, as the opposite rotation of $\delta A_{z}$,
$\delta\phi_{z}$ under the action of the complex structure (e.g.
Kapustin and Witten $^{[2]}$ (4.3))

The unique \textquotedblleft normalizable\textquotedblright\ solution is:%
\begin{align}
F(\mathbbm{r})  &  =\frac{\mathbbm{r}}{\mathtt{sh}\mathbbm{r}},\\
G(\mathbbm{r})  &  =\frac{1}{\mathbbm{r}}-\mu\mathtt{cth}\mathbbm{r}.
\end{align}
This solution is the unique solution such that the field strength has
appropriate asymptotical property both at the infinity and at the origin. When
we calculate the left(right) part of our solution, we keep $\bar{r}_{-}%
$($r_{+}$) fixed. This implies as in section 4, push down to the worldsheet by
the static (with respect to "time" $r_{-}(\bar{r}_{+})$) gauge, the dependence
on worldsheet coordinates $z=r_{-}(\bar{z}=\bar{r}_{+})$ turns to be
(anti)holomorphic, respectively.

From the outgoing comoving null frames( equations (\ref{25})-(\ref{31})), the
antiholomorphic SD field strength equals the Hermitian conjugate of ASD part,
only the sign of $\bar{H}_{T^{+}},\bar{H}_{T^{-}}$ changes, since the
orientation of the basis is reverse.

\bigskip

$\mathfrak{\bar{e}}^{T^{+}}=-\frac{1}{\sqrt{2}}e^{-i\bar{\gamma}%
}\mathbbm{r}\left(  d\bar{\theta}-\mathrm{i}\sin\bar{\theta}d\bar{\varphi
}\right)  $,$\mathfrak{\bar{e}}^{r^{\pm}}=\frac{1}{\sqrt{2}}\bar{\mu}^{\mp
1}(d\bar{r}\pm d\bar{t}),$ $\mathfrak{\bar{e}}^{T^{-}}=\frac{1}{\sqrt{2}%
}e^{i\bar{\gamma}}\mathbbm{r}\left(  d\bar{\theta}+\mathrm{i}\sin\bar{\theta
}d\bar{\varphi}\right)  $.

\section{The Dirac equation in the affine monopole background}

Now we turn to the Dirac equation in this affine monopole background:
\begin{align}
\text{For the right D bundle \ \ }\gamma^{\mu}D_{\mu}|\psi\rangle &
=0,\label{incoming Dirac equaction}\\
\text{For the left D bundle }\langle\psi|\left(  \gamma^{\bar{\mu}}D_{\bar
{\mu}}\right)  ^{\dagger}  &  =0.
\end{align}

As in last section(c.f. appendix B), we decompose $A=A_{\mu}^{\hat{a}%
}\mathbf{T}^{\hat{a}}dx^{\mu}$ into the potential $H_{\mu}$ of Dirac monopole
and the vector boson $K_{\mu}$ covariant with respect to both the spin $S$ and
the \textquotedblleft$R$-spin\textquotedblright\ $T.$
\begin{equation}
A_{\mu}=H_{\mu}+K_{\mu}.
\end{equation}
Let%
\[
D_{\mu}=D_{\mu}^{(H)}+K_{\mu}.
\]
For the right D module, the outgoing spherical waves propagate along $r^{+}$
with $r^{-}=constant$. We introduce
\begin{equation}
\kappa\equiv-i\epsilon_{ijk}\sigma_{i}\hat{r}_{j}D_{k}^{(H)}+1.
\end{equation}
here $i,j,k=T^{+},T^{-},r^{+}$, then one can proof that
\begin{equation}
\gamma_{i}D_{i}^{(H)}=\gamma_{r^{+}}\left(  {\frac{\partial}{\partial r^{+}}%
}+{\frac{1}{r^{+}}}\right)  -{\frac{1}{r^{+}}}\gamma_{r^{+}}\kappa,
\label{kappa operator}%
\end{equation}
where $\gamma_{r^{+}}=\Sigma^{2}\otimes\sigma_{r^{+}}$, $\sigma_{r^{+}%
}=\underset{i}{\sum}\sigma_{i}\hat{r}_{i}^{+}.$

Let the fixed frame wave function in the tensor product form of
\ \textquotedblleft$R$-spin\textquotedblright\ $\left\vert I\right\rangle
_{\nu}^{\beta}$ and the space time spin $\left\vert S\right\rangle _{\lambda
}^{\alpha}$ be factorized into $(t,r)$ and $(\varphi,\theta)$ dependent
part\footnote{we adopt the convention $\left\vert v\right\rangle =\left(
\begin{array}
[c]{c}%
v_{1}\\
\vdots\\
v_{n}%
\end{array}
\right)  $=$\underset{i=1}{\overset{n}{\sum}}v_{i}\left\vert e\right\rangle
_{i},\qquad\left\vert e\right\rangle _{1}=\left(
\begin{array}
[c]{c}%
1\\
0\\
\vdots\\
0
\end{array}
\right)  ,\ldots.$}%
\begin{equation}
|\psi\rangle=\psi_{\lambda\nu}^{\alpha\beta}\left\vert S\right\rangle
_{\lambda}^{\alpha}\left\vert I\right\rangle _{\nu}^{\beta},\quad\quad
\alpha,\beta=+,-;\qquad\lambda,\nu=\pm\frac{1}{2}. \label{47}%
\end{equation}

Then the incoming Dirac equation (\ref{incoming Dirac equaction}) becomes
\begin{align}
\gamma_{\mu}D_{\mu}^{(H)}  &  \left\vert \psi\right\rangle +\underset
{a,j=1,2}{\sum}\epsilon_{r^{+}aj}(\Sigma^{2}\otimes\sigma_{j})_{\lambda
\lambda^{\prime}}^{\alpha\alpha^{\prime}}\mathbb{F}(\mathbbm{r})_{\lambda
^{\prime}\lambda^{\prime\prime};\nu^{\prime}\nu^{\prime\prime}}^{\alpha
^{\prime}\alpha^{\prime\prime};\beta^{\prime}\beta^{\prime\prime}}(\Sigma
^{2}\otimes\sigma_{a})_{\nu\nu^{\prime}}^{\beta\beta^{\prime}}\psi
_{\lambda^{\prime\prime}\nu^{\prime\prime}}^{\alpha^{\prime\prime}%
\beta^{\prime\prime}}\left\vert S\right\rangle _{\lambda^{\prime\prime}%
}^{\alpha^{\prime\prime}}\left\vert I\right\rangle _{\nu^{\prime\prime}%
}^{\beta^{\prime\prime}}\nonumber\\
&  +\eta_{r^{-}r^{+}}(\Sigma^{r^{+}}\otimes\sigma_{r^{-}})_{\lambda
\lambda^{\prime}}^{\alpha\alpha^{\prime}}\mathbb{G}(\mathbbm{r})_{\lambda
^{\prime}\lambda^{\prime\prime};\nu^{\prime}\nu^{\prime\prime}}^{\alpha
^{\prime}\alpha^{\prime\prime};\beta^{\prime}\beta^{\prime\prime}}%
(\Sigma^{r^{-}}\otimes\sigma_{r^{+}})_{\nu\nu^{\prime}}^{\beta\beta^{\prime}%
}\psi_{\lambda^{\prime\prime}\nu^{\prime\prime}}^{\alpha^{\prime\prime}%
\beta^{\prime\prime}}\left\vert S\right\rangle _{\lambda^{\prime\prime}%
}^{\alpha^{\prime\prime}}\left\vert I\right\rangle _{\nu^{\prime\prime}%
}^{\beta^{\prime\prime}}=0, \label{Dirac}%
\end{align}
where
\begin{equation}
\mathbb{F}_{\lambda\lambda^{\prime};\nu\nu^{\prime}}^{\alpha\alpha^{\prime
};\beta\beta^{\prime}}\equiv\left(  (\gamma_{+})_{\lambda\lambda^{\prime}%
}^{\alpha\alpha^{\prime}}\underset{ST}{\otimes}(\rho)_{\nu\nu^{\prime}}%
^{\beta\beta^{\prime}}\right)  \frac{F(\mathbbm{r})}{\mathbbm{r}}%
,\quad\mathbb{G}_{\lambda\lambda^{\prime};\nu\nu^{\prime}}^{\alpha
\alpha^{\prime};\beta\beta^{\prime}}\equiv\left(  (\gamma_{0})_{\lambda
\lambda^{\prime}}^{\alpha\alpha^{\prime}}\underset{ST}{\otimes}(\rho)_{\nu
\nu^{\prime}}^{\beta\beta^{\prime}}\right)  G(\mathbbm{r}).;
\end{equation}
and the $\gamma_{\mu}$ matrices is decomposed as the direct product $\otimes$
of the $4d$ chirality $\sum$ and the spin $\sigma,$%
\[
\gamma_{i}=\Sigma^{2}\otimes\sigma_{i},\gamma_{0}=\Sigma^{1}\otimes\sigma
_{0,}\gamma_{5}=\Sigma^{3}\otimes\mathbf{1,}%
\]%
\begin{equation}
\sigma_{1}=\left(
\begin{array}
[c]{cc}%
0 & 1\\
1 & 0
\end{array}
\right)  ,\sigma_{2}=\left(
\begin{array}
[c]{cc}%
0 & -i\\
i & 0
\end{array}
\right)  , \label{sigma1}%
\end{equation}%
\begin{equation}
\sigma_{r^{+}}=\sigma_{3}=\left(
\begin{array}
[c]{cc}%
1 & 0\\
0 & -1
\end{array}
\right)  ,\sigma_{r^{-}}=\sigma_{0}=-\left(
\begin{array}
[c]{cc}%
i & 0\\
0 & i
\end{array}
\right)  , \label{sigma2}%
\end{equation}%
\begin{equation}
\Sigma^{r^{+}}\equiv\Sigma^{1}=\left(
\begin{array}
[c]{cc}%
0 & 1\\
1 & 0
\end{array}
\right)  ,\Sigma^{r^{-}}\equiv\Sigma^{2}=\left(
\begin{array}
[c]{cc}%
0 & -i\\
i & 0
\end{array}
\right)  ,\Gamma\equiv\Sigma^{3}=\left(
\begin{array}
[c]{cc}%
1 & 0\\
0 & -1
\end{array}
\right)  ,\Gamma^{\pm}=\frac{1}{2}(1\pm\Gamma).
\end{equation}
and $\eta^{r_{-}r_{+}}\ $is the light cone metric in real null direction.

We use the $D$ function to transform from the fixed frame basis $\left\vert
S\right\rangle =\left(
\begin{array}
[c]{c}%
\left\vert S\right\rangle ^{+}\\
\left\vert S\right\rangle ^{-}%
\end{array}
\right)  ,\left\vert I\right\rangle =\left(
\begin{array}
[c]{c}%
\left\vert I\right\rangle ^{+}\\
\left\vert I\right\rangle ^{-}%
\end{array}
\right)  $ to the comoving frame $|s>_{\lambda}^{\pm},|i>_{\nu}^{\pm}$%
\begin{equation}
{}|s>_{\lambda}^{\pm}=\sum_{\rho}D_{\rho\lambda}^{\pm S}(\varphi,\theta
,\gamma)|S>_{\rho}^{\pm},\left\vert S\right\rangle _{\frac{1}{2}}=\left(
\begin{array}
[c]{c}%
1\\
0
\end{array}
\right)  ,\left\vert S\right\rangle _{-{\frac{1}{2}}}=\left(
\begin{array}
[c]{c}%
0\\
1
\end{array}
\right)  ,
\end{equation}%
\begin{equation}
{}|i>_{\lambda}^{\pm}=\sum_{\rho}D_{\rho\lambda}^{\pm T}(\varphi,\theta
,\gamma)|I>_{\rho},\left\vert I\right\rangle _{\frac{1}{2}}=\left(
\begin{array}
[c]{c}%
1\\
0
\end{array}
\right)  ,\left\vert I\right\rangle _{-{\frac{1}{2}}}=\left(
\begin{array}
[c]{c}%
0\\
1
\end{array}
\right)  , \label{after 56}%
\end{equation}
which satisfy for the spin part: $\sigma_{r}|s>_{\lambda}^{\pm}=2\lambda
|s>_{\lambda}^{\pm}$, $\sigma_{r}|i>_{\nu}^{\pm}=2\nu|i>_{\nu}^{\pm},$ for the
chirality part: $\Gamma|i>^{\pm}=\pm|i>^{\pm},\Gamma|s>^{\pm}=\pm|s>^{\pm}.$
The superscript "$S$" and "$T$" denote the finite rotation matrix function
$D_{\lambda\rho}^{\frac{1}{2}}(\varphi,\theta,\gamma)$ act in the spin space
and the isospin space, respectively, with $S={\frac{1}{2}},T={\frac{1}{2}}.$

Then the fixed frame eq. (\ref{47}) (\ref{Dirac}) is transformed to the
following comoving frame eq.%

\begin{align}
&  \left(  \gamma_{r^{+}}({\frac{\partial}{\partial r^{+}}}+{\frac{1}{r^{+}}%
})\left\vert \psi\right\rangle \right) \label{radial}\\
&  +\{\underset{a,j=1,2}{\sum}\epsilon_{3aj}(\hat{\Sigma}^{2}(x)^{\alpha
\alpha^{\prime}}\otimes\hat{\sigma}_{j}(x)_{\lambda\lambda^{\prime}%
})(D_{\lambda^{\prime\prime}\lambda^{\prime}}^{\alpha S})\underset
{ST}{{\otimes}}\mathbb{F}_{\lambda^{\prime\prime}\lambda^{\prime\prime\prime
};\nu^{\prime\prime}\nu^{\prime\prime\prime}}^{\alpha^{\prime}\beta^{\prime
};\alpha^{\prime\prime}\beta^{\prime\prime}}(\mathbbm{r})(D_{\nu^{\prime
\prime}\nu^{\prime}}^{\beta T})(\hat{\Sigma}^{2}(x)^{\beta\beta^{\prime}%
}\otimes\hat{\sigma}_{a}(x)_{\nu\nu^{\prime}})\nonumber\\
&  +(\hat{\Sigma}^{r^{+}}(x)^{\alpha\alpha^{\prime}}\otimes\hat{\sigma}%
_{r^{-}}(x)_{\lambda\lambda^{\prime}})(D_{\lambda^{\prime\prime}%
\lambda^{\prime}}^{\alpha S})\eta_{r^{-}r^{+}}\underset{ST}{{\otimes}}%
(D_{\nu^{\prime\prime}\nu^{\prime}}^{\beta T})\mathbb{G}_{\lambda
^{\prime\prime}\lambda^{\prime\prime\prime};\nu^{\prime\prime}\nu
^{\prime\prime\prime}}^{\alpha^{\prime}\beta^{\prime};\alpha^{\prime\prime
}\beta^{\prime\prime}}(\mathbbm{r})(\hat{\Sigma}^{r^{-}}(x)^{\beta
\beta^{\prime}}\otimes\hat{\sigma}_{r^{+}}(x)_{\nu\nu^{\prime}})\}\nonumber\\
&  \times f_{\lambda^{\prime\prime\prime}\nu^{\prime\prime\prime}}%
^{\alpha^{\prime\prime}\beta^{\prime\prime}}\left\vert s\right\rangle
_{\lambda^{\prime\prime\prime}}^{\alpha^{\prime\prime}}\left\vert
i\right\rangle _{\nu^{\prime\prime\prime}}^{\beta^{\prime\prime}}=0,
\end{align}

\begin{equation}
\hat{\psi}_{\lambda\nu}^{\alpha\beta}=f_{\lambda\nu}^{\alpha\beta
}(r,t)\mathbb{D}_{\lambda\nu}^{\alpha\beta}(\varphi,\theta,\gamma).
\end{equation}

Here the $\hat{\sigma}(x),\hat{\Sigma}(x)$ matrix acted on the comoving frame
basis $|i>_{\nu}^{\pm},$ $|s>_{\lambda}^{\pm}$becomes the cons$\tan$t matrices
in (\ref{sigma1})(\ref{sigma2}) (c.f. (B8)-(B12)). Then, since the $\kappa$
operator in (\ref{kappa operator}) turns to be zero as shown after (\ref{15})
, similar as (B13), this Dirac equation turns to be simply a "tangent vector
equation along the left radial direction"

We note that in the second term of the eq. (\ref{radial}) the space spin $S$
and the $R$ spin $T$ are coupled into
\begin{align}
&  \epsilon_{3aj}(\sigma_{j})_{\lambda^{\prime}\lambda^{\prime\prime}%
}D_{\lambda\lambda^{\prime}}^{\alpha S}\left\vert S\right\rangle
_{\lambda^{\prime\prime}}^{\alpha}\underset{ST}{{\otimes}}(\sigma_{a}%
)_{\nu^{\prime}\nu^{\prime\prime}}D_{\nu\nu^{\prime}}^{\beta T}\left\vert
I\right\rangle _{\nu^{\prime\prime}}^{\beta}\nonumber\\
&  =\epsilon_{3aj}{}(\hat{\sigma}_{j})_{\lambda\lambda^{\prime}}\,\left\vert
s\right\rangle _{\lambda^{\prime}}^{\alpha}\underset{ST}{{\otimes}}%
(\hat{\sigma}_{a})_{\nu\nu^{\prime}}\left\vert i\right\rangle _{\nu^{\prime}%
}^{\beta}(-1)^{\lambda-\frac{1}{2}}(-\delta_{\alpha,\beta}\delta_{\lambda,\nu
}+\delta_{\alpha,-\beta}\delta_{\lambda,-\nu}), \label{3.14}%
\end{align}

\bigskip with nonvanishing $(\varphi,\theta)$ functions $\mathbb{D}$
\begin{equation}
\mathbb{D}_{\lambda\nu}^{\mathbb{\alpha\beta}}(\varphi,\theta,\gamma
)\equiv(-1)^{2\lambda+1}(-\delta_{\lambda,\nu}\delta_{\alpha,\beta}%
+\delta_{\lambda,-\nu}\delta_{\alpha,-\beta})\left\vert s\right\rangle
_{\lambda}^{\alpha}\left\vert i\right\rangle _{\nu}^{\beta}.
\end{equation}
by using
\begin{align}
&  \frac{1}{4\pi}\underset{\lambda^{\prime},\nu^{\prime}}{\sum}C_{\lambda
^{\prime},\nu^{\prime},0}^{S,\,T,\,0}D_{\lambda^{\prime}\lambda}^{\alpha
S}(\varphi,\theta,\gamma)D_{\nu^{\prime}\nu}^{-\alpha T}(\varphi,\theta
,\gamma)=\frac{1}{\sqrt{2\pi}}D_{0,\lambda+\nu}^{0}(\varphi,\theta
,\gamma)C_{\lambda,\nu,0}^{S,T,0}=(-1)^{\lambda-\frac{1}{2}}\delta
_{\lambda,-\nu},\\
&  \frac{1}{4\pi}\underset{\lambda^{\prime},\nu^{\prime}}{\sum}C_{\lambda
^{\prime},\nu^{\prime},0}^{S,\,T,\,0}D_{\lambda^{\prime}\lambda}^{\alpha
S}(\varphi,\theta,\gamma)D_{\nu^{\prime}\nu}^{\alpha T}(\varphi,\theta
,\gamma)=\frac{1}{\sqrt{2\pi}}D_{0,\lambda-\nu}^{0}(\varphi,\theta
,\gamma)C_{\lambda,-\nu,0}^{S,\,T,0}=-(-1)^{\lambda-\frac{1}{2}}%
\delta_{\lambda,\nu}, \label{15}%
\end{align}
here the finite rotation matrices $D_{\mu^{\prime}\mu}^{S}$ and $D_{\nu
^{\prime}\nu}^{T}$ from the fixed frames to the comoving frames is coupled to
a singlet expressed by the well known Kroneker matrix $(-1)^{\mu-\frac{1}{2}%
}\delta_{\mu,\mp\nu}$, and we use $\epsilon^{3aj}$ as the $C_{\lambda
\,\,\,\nu\,\,0}^{\frac{1}{2}\frac{1}{2}0}$ to couple $D^{S}$ and $D^{T}$ to
obtain $D_{m,\mu+\nu}^{J}s_{\mu}i_{\nu}$, this gives $J=0$, $\delta_{\mu
,\pm\nu}$ and so $\kappa=0$ in eq.~(\ref{radial}) implicitly. Since the $S$
$T$ chirality have opposite helicity, so there are two terms in (\ref{3.14}),
with opposite sign.

We also note that for the last term of the eq. (\ref{radial}), $S$ and $T$ are
coupled as
\begin{equation}
\eta_{^{r^{-}r^{+}}}\sigma_{^{r^{-}}}D^{\alpha S}\left\vert S\right\rangle
\otimes\sigma_{r^{+}}D^{\beta T}\left\vert I\right\rangle =\eta^{r^{-}r^{+}%
}(\sigma^{r^{-}}{}|s>_{\lambda}^{\alpha}\underset{ST}{{\otimes}}\sigma^{r^{+}%
}\left\vert i\right\rangle _{\nu}^{\beta})(-1)^{\lambda-\frac{1}{2}}%
(-\delta_{\alpha,\beta}\delta_{\lambda,\nu}+\delta_{\alpha,-\beta}%
\delta_{\lambda,-\nu}),
\end{equation}
Thus, the $D$ funtion has been simply coupled into the Kronecker matrix.

From constraint for the zero mode sections in (\ref{after 56}), the original
$16\otimes16$ tensor product space $\left\vert s\right\rangle _{\lambda
}^{\alpha}\left\vert i\right\rangle _{\nu}^{\beta}$ is reduced to $8$
dimension, including $\left\vert s\right\rangle _{\lambda}^{\alpha}\left\vert
i\right\rangle _{\lambda}^{\alpha}$ and $\left\vert s\right\rangle _{\lambda
}^{\alpha}\left\vert i\right\rangle _{-\lambda}^{-\alpha}$ only. Then \ after
factorize out the $\varphi,\theta$ dependent basis $\left\vert i\right\rangle
,$ $\left\vert s\right\rangle ,$ we can show as (B16)-(B18) that since for the
zero mode section we have further constraint $f_{\lambda\nu}^{\alpha\beta
}=-f_{-\lambda-\nu}^{\alpha\beta}$ $\equiv f^{\alpha\beta}.$ So only 4
independent component remain only.

Remind that, as has been proved by Hitchin$^{[29]}$ only the left
imaginary null $\mathbf{K}_{T^{-}}$ and real null
$\mathbf{K}_{r^{-}}$ contribute on $\alpha$ plane. The Dirac
equation in the cup product frame turns to be
\begin{equation}
\left(  ({\frac{\partial}{\partial r^{+}}}+{\frac{1}{r^{+}}})I+\mathbf{K}%
_{T^{-}}+\mathbf{K}_{r^{-}}\right)  \left\vert \xi\right\rangle \equiv\left(
({\frac{\partial}{\partial r^{+}}}+{\frac{1}{r^{+}}})I+\mathbb{K}_{+}\right)
\left\vert \xi\right\rangle =0, \label{imp}%
\end{equation}%
\[
\left\vert \xi\right\rangle \equiv\left\vert
\begin{array}
[c]{c}%
f^{++}\\
-f^{+-}\\
f^{--}\\
-f^{-+}%
\end{array}
\right\rangle ,
\]
where $\mathbf{K}_{\mu}$ are%
\begin{align*}
\mathbf{K}_{T^{-}}  &  =\underset{i=1}{\overset{4}{\sum}}\frac{F^{i}%
(\mathbbm{r})}{\mathbbm{r}}\mathbf{E}_{i,i+1},\quad\mathbf{K}_{T^{+}%
}=-\underset{i=1}{\overset{4}{\sum}}\frac{F^{i}(\mathbbm{r})}{\mathbbm{r}}%
\mathbf{E}_{i+1,i}\\
\mathbf{K}_{r^{-}}  &  =\frac{1}{2}\underset{i=1}{\overset{4}{\sum}}%
G^{i}(\mathbbm{r})(\mathbf{E}_{i,i}+\mathbf{E}_{i+1,i+1}),\quad\\
F^{i}(\mathbbm{r})  &  =\omega^{i}F(\mathbbm{r}),\quad G^{i}%
(\mathbbm{r})=\omega^{i}G(\mathbbm{r}),
\end{align*}
here we add an subscript $+$ to $\mathbb{K=}\mathbf{K}_{T^{-}}+\mathbf{K}%
_{r^{-}},$ to denote that it is the whole $\mathbb{K}$ along $r^{+}$. Further,
we write $\mathbb{A}_{+}$ instead $\mathbf{K}_{+},$ since the $D_{\mu}^{(H)}$
becomes simply ${\frac{\partial}{\partial r}}+{\frac{1}{r}}$ in this gauge, by
the vanishing of $\kappa.$ The eq. (\ref{imp}) in each $i$-th $2\otimes2$
block generated by the $\mathbf{E}_{i+1,i}$, $\mathbf{E}_{i,i+1}$and
$\mathbf{E}_{i,i}$, $\mathbf{E}_{i+1,i+1}$ is obtained in the same way as
appendix B. Here we generalize the cup product $\underline{2}\otimes
\underline{2}\rightarrow\underline{1}\otimes$\b{3} into $(2,1)\otimes
((2,1)\oplus(1,2))\rightarrow(1,1)\oplus(3,1)$ and conjugate$.$ That is the
same decomposition of the quaternion product for the kernel of Dirac operator
in SDYM. The outgoing Dirac equation is manipulated in the same way.%
\begin{equation}
\left\langle \xi\right\vert \left(  ({\frac{\partial}{\partial\bar{r}^{-}}%
}+{\frac{1}{\bar{r}^{-}})I+}\mathbb{K}_{-}\right)  =0
\end{equation}
For the consistence of this equation with (\ref{imp}) at the intersection
points of the outgoing and incoming waves, we should affinize the algebra by
the inclusion of the $\mathbf{c}$ and the $\mathbf{d}$ term (see next
section). The explicit form of $f^{\alpha\beta}(r^{+},r^{-})$ will be find in
next section by solving the scattering equation of the incoming and outgoing waves.

\section{Conformal affine massive models, affine ADHMN construction and 't
Hooft Hecke operator}

In this section we will sketch how this zero mode solutions in the cup product
form (section 3) in the background of the flat connection of the D bundles
(section 2) are transformed to the conformal \textbf{affine} system by the
affine Nahm construction, then turn to the 't Hooft Hecke operator.

In previous section, by transform from the fixed frame to the tensor product
form of the spherical comoving frame, the dependence of $\theta,\varphi
,\gamma(\theta,\varphi)$ disappear by \textquotedblleft Fourier
transformation\textquotedblright, i.e. integrate with the Green function for
these variable. Meanwhile we have choose the geodesics coordinate $\gamma
=\pm\varphi$ in the north (south) patch of $S^{2}$ ($\sim\mathbb{C}P_{1})$.
But the dependence on $r^{+}$ $r^{-}$ remains, so we call it intermediate
comoving frame. Now to get the final comoving frame without $r^{+}$ $r^{-}$
dependence. we adopt the following step. We start from the cup product form in
(\ref{imp}), then, firstly, we rotate the $\left\vert \xi\right\rangle $ in
(\ref{imp}) by an angle $\phi$ such that $e^{\frac{1}{2}\phi}%
=coth(\mathbbm{r})$ (c.f(\ref{kk})) , so that $D_{r^{+}}$ in (\ref{imp}) for
each $G^{i},$ $F^{i}$ term of the cup product becomes $\left(  \frac{\partial
}{\partial\mathbbm{r}}+\frac{F^{i}\mathbbm{r})}{\mathbbm{r}}\left(
\begin{array}
[c]{cc}%
0 & 1\\
1 & 0
\end{array}
\right)  +G^{i}(\mathbbm{r})-\frac{1}{\mathbbm{r}}\right)  .$ By noticing
that$\left(  \frac{d}{d\mathbbm{r}}+\left(  G^{i}(\mathbbm{r})-\frac
{1}{\mathbbm{r}}\right)  \right)  \frac{F^{i}(\mathbbm{r})}{\mathbbm{r}}=0$,
i.e. the diagonal function are the logarithm derivative of the off diagonal
function. In these geodesic $r,t$ fixed frame, we have $\mathbb{A}_{\pm}%
^{i}=\left(
\begin{array}
[c]{cc}%
G^{i}(\mathbbm{r})-\frac{1}{\mathbbm{r}} & \frac{F^{i}(\mathbbm{r})}%
{\mathbbm{r}}\\
\frac{F^{i}(\mathbbm{r})}{\mathbbm{r}} & G^{i}(\mathbbm{r})-\frac
{1}{\mathbbm{r}}%
\end{array}
\right)  $ $\pm\zeta C.$ Then for the outgoing and incoming waves scattering
along $r_{+},$ $r_{-}$ direction. The fixed frame $\mathbb{A}_{\pm}$ is given
by integrate the light cone comoving frame $a_{\pm}$ along the spectral line.

\noindent%
\[
\mathbb{A}_{\pm}=Pexp\int a_{\pm}(\mathbbm{z};\mu),
\]

\noindent where
\begin{equation}
a_{\pm}(\mathbbm{z};\mu)=a_{\pm}(\mathbbm{z})=\frac{\partial_{z_{\pm}%
}\Big(G-\frac{1}{\mathbbm{z}}\Big)}{\Big(G-\frac{1}{\mathbbm{z}}\Big)}%
\rho^{\mp2}+\Big(\frac{\partial_{z_{\pm}}\big(\frac{F}{\mathbbm{z}}%
\big)}{\big(\frac{F}{\mathbbm{z}}\big)}\Big)^{\pm1}\rho^{\mp1}E^{\pm1}%
+\zeta(\mathbbm{z})c \label{80}%
\end{equation}
here we have push down $\mathbbm{r}=\mu r_{+}+\mu^{-1}\bar{r}_{-}$ to
$z_{+}=r_{+},z_{-}=r_{-}$ on worldsheet. This $a_{\pm}(\mathbbm{z})$turns to
be the Lax connection$\mathbb{\ }a_{\pm}(\mathbbm{z})=\underset{i=0}%
{\overset{3}{\sum}}$ $\partial_{z_{\pm}}\phi_{i}\omega^{\pm2i}E_{i,i}%
+e^{\pm\phi_{i}}(E^{\pm1})_{i,i\pm1}$ , as we substitute in it the one soliton
solution of the conformal affine Toda eq., by the following identification%
\begin{align}
\frac{\left(  \frac{F(\mathbbm{z})}{\mathbbm{z}}\right)  ^{\prime}}%
{\frac{F(\mathbbm{z})}{\mathbbm{z}}}  &  =e^{\left\vert \phi_{soliton}%
\right\vert }=\frac{1+e^{2\mathbbm{z}}}{1-e^{2\mathbbm{z}}}=\coth
\mathbbm{z},\ \phi_{i}=\left\vert \phi\right\vert \rho_{i},\quad
\quad\nonumber\\
\eta &  =\frac{\pi\mathrm{i}}{2},d\eta=0,\zeta=\ln(1-e^{2\mathbbm{z}}%
)\equiv\zeta_{sol}-\zeta_{vac}. \label{kk}%
\end{align}

The Lax equation implies the existence of the transport matrix $U,$
$\partial_{\pm}U=a_{\pm}(\mathbbm{z})U.$ Let the monodromy matrix
$\mathcal{T}_{\pm}$, the loop operator of $a$ along $r_{\pm}=0\rightarrow
\infty,$ $\mathcal{T_{+}}=\int_{0}^{\infty}a(\mathbbm{z})dz$, $\mathcal{T_{-}%
}=\int_{0}^{\infty}a(\mathbbm{z})d\bar{z},$ then $\mathcal{T_{\pm}}$ becomes
independent of $r,t$ and satisfy the \textbf{affine Nahm eq. or the affine
Donaldson's imaginary eq.}
\begin{align}
\lambda\frac{d\mathcal{T}_{+}^{\alpha}(\lambda)}{d\lambda}-[\mathcal{T}%
_{+}^{h}(\lambda),\mathcal{T}_{+}^{\alpha}(\lambda)]-\delta_{+}(s) &  =0;\\
-\lambda\frac{d\mathcal{T}_{-}^{\alpha}(\lambda)}{d\lambda}-[\mathcal{T}%
_{-}^{h}(\lambda),\mathcal{T}_{-}^{\alpha}(\lambda)]-\delta_{-}(s) &  =0;
\end{align}
\textbf{real} eq.
\begin{equation}
\lambda\frac{d\mathcal{T}_{+}^{\mathfrak{h}}(\lambda)}{d\lambda}%
-\frac{d\mathcal{T}_{-}^{\mathfrak{h}}(\lambda)}{d\lambda}+[\mathcal{T}%
_{+}^{\mathfrak{h}}(\lambda),\mathcal{T}_{-}^{\mathfrak{h}}({\lambda
})]+[\mathcal{T}_{+}^{\alpha}(\lambda),\mathcal{T}_{-}^{\alpha}(\lambda
)]-\delta(s)=\zeta(\mathbbm{r}_{0}).
\end{equation}
which depend on $\lambda=e^{s}$ only, here $\lambda$ is the loop parameter of
$\hat{g}$ and $\tilde{g}$, $\mathcal{T}_{\pm}(\lambda)=\mathcal{T}%
_{-}^{\mathfrak{h}}({\lambda})\rho+\mathcal{T}_{\pm}^{\mathfrak{\pm}}%
({\lambda})E^{\pm1}$
\[
\delta_{\pm}(s)=\frac{1}{2}(\delta(s)\pm\frac{i}{\pi}\mathcal{P}\frac{1}{s}).
\]

Meanwhile we change the $r_{+},$ $r_{-}$ with respect to the scattering point
(blow up point) $\mathbbm{r}_{0}$ by $r_{+}\rightarrow r_{+}-r_{0+}$,
$r_{-}\rightarrow r_{-}-r_{0-},$ $\mathbbm{r}_{0}\equiv\mu r_{0+}+\bar{\mu
}r_{0-}.$

\bigskip The residue of $\mathcal{T}$ around $\lambda=\mu,\bar{\lambda}%
=\bar{\mu}$ respectively are the $E$ $E^{-1}$ generators of the $A_{3}^{(1)}$.
The parameter $\mu$ describes the central $U(1)^{c}$ of $U(4),$ i.e. the
diffeomorphism, area preserving, dual twistor angle $U(1)$ symmetry, which we
affinize, and it happens that this serves also as the common dilation,
rapidity shift parameter in Rindler coordinate. This dilation operators also
rotate the fixed frame $\left\vert \Lambda_{\max}\right\rangle $ to the moving
frame $|\xi_{vac}\rangle.$ After all this we have reach the affine twistor construction.

It is easy to check that the character of the blow up sheaf $\mathcal{M}%
_{n,r,k}$ (e.g. in [21]) is the pointygin class $c_{2}=n=1$, rank
$r=4$, $1$st chern class $c_{1}=k=1$ case is the one soliton $\tau$
function of conformal affine Toda. Which can be factorize into
$\left\langle \xi (\bar{\lambda})\right\vert \left.
\xi(\lambda)\right\rangle .$

The distributions in fact are given by the trace twisted by the density matrix
$\tilde{\rho}$ (\ref{ryo1})%
\begin{align*}
Trace\left\vert \xi\right\rangle \Gamma^{\pm}\left\langle \xi\right\vert  &
=\delta_{\pm}(\lambda),\\
Trace\left\vert \xi\right\rangle \Gamma\left\langle \xi\right\vert  &
=\delta(\lambda),\\
TraceM  &  =trace\tilde{\rho}M
\end{align*}
the occurrence of the density matrix, comes from that $\left\vert
\xi\right\rangle $ as a state in massive integrable model are non-pure state.

To find the density matrix, besides the
$\lambda\frac{\partial}{\partial \lambda}\equiv d$ and $c$, one
should introduce the homotopy operator $\mathfrak{K}$. That is, we
are dealing with a massive integrable field theory\footnote{We will
following the discussion for the quantum massive integrable field by
Lukyanov in [19].} as the Unrah effect, the bare vacuum should be
replaced by introducing the density matrix $\rho$,
\begin{equation}
<\xi_{vac}|T(\lambda_{2})T(\lambda_{1})|\xi_{vac}>=tr[\tilde{\rho}%
T(\lambda_{1})T(\lambda_{2})], \label{ryo}%
\end{equation}%
\begin{equation}
\tilde{\rho}=e^{2\pi i\mathfrak{K}}, \label{ryo1}%
\end{equation}
how $\lambda$ becomes the rapidity in the Rindler coordinates:
\begin{align}
&  x=r\cosh\alpha,t=r\sinh\alpha,\nonumber\\
&  -\infty<\alpha<+\infty,0<r<+\infty.
\end{align}
We can introduce the rapidity shift operator $\mathfrak{K}$, such that,
\begin{equation}
e^{\alpha\mathfrak{K}}T(\lambda)e^{-\alpha\mathfrak{K}}=T(\lambda-\alpha).
\label{star}%
\end{equation}

where the shift operator, homotopy Bochner-Martinelli kernel operator
$\mathfrak{K}$ is realized as the second terms in the right hand side of the
following equation (\ref{BB})
\begin{align}
\lbrack\hat{X},\hat{Y}]  &  =[\tilde{X},\tilde{Y}]_{\sim}+{\frac{1}{2}}%
\oint{\frac{d\lambda}{2i\pi}}tr(\partial_{\lambda}\tilde{X}(\lambda
)\cdot\tilde{Y}(\lambda))\;C\nonumber\\
&  =[\tilde{X},\tilde{Y}]_{\sim}+\mathfrak{K}\tilde{X}(\tilde{Y})C, \label{BB}%
\end{align}
here $\hat{X},\hat{Y}\in\mathfrak{\hat{g}}$, $\tilde{X},\tilde{Y}%
\in\mathfrak{\tilde{g}.}$

Actually, the parameter $\lambda$ is introduced by following
Nahm$^{[27]}$, it is the Fourier transformation of time $t$
originally restricted to the static time of BPS monopole in [27],
but now generated to $r_{+}$, $r_{-}$ for incoming and outgoing wave
along $r_{+}$, $r_{-},$ which is further push down respectively to
$z_{+},z_{-},$ and further central extend upon scattering point
$\mathbbm{z}_{0}$. So this affine parameter is the rapidity, i.e.
the Fourier transformation of the Rindler coordinates
\textbf{$\alpha$}. Together with the rotation matrix in the base
space $D^{S}(\theta,\varphi)$ and in the t$\arg$et space
$D^{T}(\theta,\varphi)$. The radial scattering function and the
Rindler boost function expressed by the radial Bessel function gives
the Fourier transformation which constitute the affine monopole
function for an elementary blow up point.

Now as in [3], we rotate (dressing transform) the wavefunction
$|\mathcal{\xi}\rangle$ and$\langle\bar{\xi}|$ respectively from the
$|\mathcal{\xi}_{vac}\rangle$ to $|\mathcal{\xi}_{sol}\rangle$,
$\langle\bar{\xi}_{vac}|$ to $\langle\bar{\xi}_{sol}|$ at the same
time changes $a_{vac}$ to $a_{sol}$ by conjugate by the vertex
operator $V(z)$ ('t Hooft Hecke operator). The positive (negative)
frequency part $V_{-}(z)$,
$(V_{+}(z))$ is determined by the Riemann Hilbert transformation.%
\[
\tilde{V}_{-}=V_{-}e^{-\frac{1}{2}\zeta(\mathbbm{r})}%
\]%
\begin{align}
\tilde{V}_{-}^{-1}(z) &  =\frac{1}{2}\zeta^{-1}\frac{\partial}{\partial z_{+}%
}\zeta+\left(  \frac{P_{+}}{\lambda-\mu}+\frac{P_{-}}{\lambda+\mu}\right)
\nonumber\\
&  \equiv\left(
\begin{array}
[c]{cc}%
\sqrt{\frac{e^{\mathbbm{r}}+e^{-\mathbbm{r}}}{e^{\mathbbm{r}}-e^{-\mathbbm{r}}%
}I} & 0\\
0 & \sqrt{\frac{e^{\mathbbm{r}}-e^{-\mathbbm{r}}}{e^{\mathbbm{r}}%
+e^{-\mathbbm{r}}}I}%
\end{array}
\right)  +\frac{2e^{2\mathbbm{r}}}{\sqrt{1-e^{4\mathbbm{r}}}}\left(
\begin{array}
[c]{cc}%
\mu & \lambda\\
-\lambda & -\mu
\end{array}
\right)  \frac{\mu}{\lambda^{2}-\mu^{2}}\\
P_{\pm} &  \equiv\frac{\mu e^{\mathbbm{r}}}{\sqrt{1-e^{2\mathbbm{r}}}}\left(
\begin{array}
[c]{cc}%
\pm I & I\\
-I & \mp I
\end{array}
\right)  ,\label{81}%
\end{align}
here all matrices are $4\times4$, the elements in each chiral blocks are
$2\times2$ diagonal. $P_{\pm}$ is diagonal along the rotation axis for the
operator $V$ with rotation angle $\frac{1}{2}\varphi_{sol}=\coth(\mathbbm{z})$.

\textbf{Remark:} The right hand side of this R. H. eq. is similar as
the following operators in Hitchin's paper$^{[24,p51]}$
\begin{equation}
i\frac{d}{d\lambda}+(\frac{1}{\lambda-1}+\frac{1}{\lambda+1})T.
\end{equation}

The Hirota e.q. of\ $\tau$, gives the background dependent
holomorphic anomaly. After sum over various representation of
$\left\vert \xi\right\rangle $ Nakajima$^{[21]}$ obtain the quantum
$\tau$. But the perturbative factor of the character (partition
function ), has been get by conjecture to fit the large $N$
approximation. We have include the central extension with center
$\mathbf{C}$ lies in the center $\mathbb{Z}$ of $U(4)$, to construct
the universal bundle. This extension by homotopy is the way of
nonabelian localization as calculated by Beasley and Witten for the
Seifert manifold$^{[25]}$

We may match our affine Nahm eq. of the 't Hooft Hecke operator with
the Nahm eq. of the surface operator given by Gukov and
Witten$^{[26]}$, since we have both 1st and 2nd chern class $c_{1}$
and $c_{2}.$

\section{Discussion and Outlook}

In fact eq.~(\ref{BB}) is the integral kernel of the Bochner-Martinelli
formula. From this, the reside theorem will give the Poincare-Hopf
localization index, i.e. the analytic index. After doing the Thom isomorphism,
the representative of this index can be written as the integral of the A-roof
genus by using the Bott reside formula. This is just the topological index.

In fact, the analytic index is just the anti-holomorphic function on
the sheaf $T(x)$ and $\psi(x)$ can be expressed using the $T(x)$ in
[20]. The one corresponds to the real null vector, $T_{0}\pm T_{1}$,
is just $B_{1}$, and the one corresponds to the imaginary vector,
$T_{2}\pm iT_{3}$, is just $B_{2}$, and $\oint|\psi><\psi|d\lambda$
is just $i^{*}j+j^{*}i$. The character for the one affine monopole
solution is just the one in [20] with $r=2, n=4, k=1$.

If we use the Riemann-Hilbert transformation with N poles, we will
get the N-soliton solution. The corresponding character is the
character for the sheaf with the Young diagram that satisfies
$|Y|=N$ in [20].

Now we turn to the problem of quantization. First we should use the
Seiberg-Witten curve to determine the cut-off constant $\Lambda$,
here the Seiberg-Witten curve is just the spectral curve of the Lax
connections in the affine Toda system. There are four formalisms for
quantization. The first one is the quantum group formalism, such as
the massive integrable field theory (the one studied in [19]. The
bosonic oscillator representation is given in [10] and [13](here the
relevant quantum group is $\hat{Sl}_{4}(p,q)$). In the second
formalism, twistor and sheaf are used$^{[20,21]}$. The third
formalism is the field theoretic formalism, where the hyperkahler
quotient plays a important role. In the last one, the spectral curve
in the moduli space form a Calabi-Yau 3-fold. Here, in the
commutation relation, we should use $\phi_{quantum}$ to replace
the$^{[22]}$ $\phi_{vacuum}$.

In the quantum group formalism, beside expanse the $\phi_{vaccuum}$ by the
q-oscillator, we also obtain the exact explicit commutation relation of the
vertex operator. We can obtain the exact result in the thermodynamics limit
which is not an approximate result in the large N limit. Moreover, the two
kinds of vertex operators in the quantum group formalism are corresponding to
the Wilson loop and the 't Hooft loop in Yang-Mills theory, respectively.
Modular transformation will exchange both these two kinds of operators and the
corresponding states characterize respectively the order parameter and
disorder parameter. This is also called the level-rank duality.

The area-preserving transformation of the dual twist angle $\varphi$
($\lambda=e^{i\varphi}$) is just the diffeomorphism transformation. So it will
connect the phase angle $\varphi$ of the center term of the topological field
theory and the attractor parameter $\mu$ of the black hole entropy.

The quasi-Hopf quantum double of the Drinfeld's quantum group not
only realizes the almost factorization in [23], but also give the
center term of the holomorphic anomaly, which is obtained by solving
the Hirota equation. In this situation, the polarization of the
waving function can be used to explain the background independence.

\section*{Acknowledgments}

The author would like to thank for the support of National Natural
Science Foundation of China under Grant No.90403019. we would like
to thank X. C. Song, K. J. Shi, K. Wu, S. Hu, Y. X.Chen, Y. Z.
Zhang, R. H. Yue, L. Zhao, X. M. Ding, C. H. Xiong, X. H. Wang and
S. M. Ke for their helpful discussion and would like to thank J. B.
Wu for helpful discussions and for the help to write the preliminary
version of this paper; thanks Simonsyang and binlijunwang for their
helpful discussions on the version 1 and version 2 of this paper.

\section*{Appendix A. BPS monopole}

The static spherical symmetric ansatz \footnote{For the BPS monopole which is
constituted by the Higgs Yang-Mills field or spontaneously breaking Yang-Mills
field, the $A_{0}^{a}$ should be replaced by the Higgs field $i\Phi^{a\text{
}}$and the ASD equation by the BPS equation.
\par
Thus, we adopt the notation (A2) in view of further embedding into the SUSY
affine monopole as the incoming and outgoing waves in the Sec. 2 and Sec. 3.
There, in case of $d=4$ SUYM, our ansatz $A_{\mu}$ turns to be the
$\mathcal{A}_{\mu}$ of paper [2], $\mathcal{A}_{\mu}%
=A_{\mu}+i\phi_{\mu+4}$ with $A_{0}=\phi_{5}=\phi_{6}=\phi_{7}=0$.}
\begin{align}
A_{i}^{a}(x)  &  =\epsilon_{iaj}\frac{F(r)-1}{r}\hat{r}^{j},\tag{A1}\\
\qquad A_{0}^{a}(x)  &  =iG(r)\hat{r}^{a}(x),\qquad\tag{A2}%
\end{align}
where $r=(x_{1}^{2}+x_{2}^{2}+x_{3}^{2})^{\frac{1}{2}}$, the unit radial
vector $\hat{r}^{j}=\frac{r^{j}}{r},$ the space spin index $\ i,j=1,2,3$, the
"isospin" $SU(2)$ generator index $a=1,2,3$. This ansatz is $U(1)$ symmetrical
under the cooperative local rotation generated by both the space time spin
$\hat{s}(x)=\hat{r}$ and the $SU(2)$ isospin $\mathbf{T}^{r}(x)=\underset
{a}{\sum}T^{a}\hat{r}^{a}$.

Decompose $A$ as
\begin{equation}
A_{\mu}=H_{\mu}+K_{\mu},H_{i}^{a}=-\varepsilon_{iaj}\frac{1}{r}\hat{r}%
^{j},K_{i}^{a}=\epsilon_{iaj}\frac{F(r)}{r}\hat{r}^{j},K_{0}^{a}=iG(r)\hat
{r}^{a},H_{0}^{a}=0. \tag{A3}%
\end{equation}
Let\footnote{In appendices, we adopt the physics convention with a Hermitian
gauge field $A_{\mu}.$} $D_{\mu}=\partial_{\mu}-iA_{\mu}=D_{\mu}^{(H)}%
-iK_{\mu},$

\noindent herein
\begin{equation}
D_{\mu}^{(H)}=\partial_{\mu}-iH_{\mu}. \tag{A4}%
\end{equation}

Then we have that, $\mathbf{T}^{r}(x)$ is covariant constant under the $SU(2)
$ "isospin" rotation; $D_{i}^{(H)}\mathbf{T}^{r}(x)=(\partial_{i}+iH^{a}%
_{i}T^{[a})(\sum_{b}T^{b]}\hat{r}^{b})=\partial_{i}\hat{r}^{a}T^{a}
-i\epsilon_{iaj}\hat{r}^{j}\hat{r}^{b}[T^{a},T^{b}]=0$; $D_{0}^{(H)}%
\mathbf{T}^{r}(x)=0$; and meanwhile $\nabla_{i}\hat{r}^{a}=0$ under the action
of spin connection along the surface $S^{2}$ ($r=constant)$.

Transform from the spacetime fixed frames in \textquotedblright spherical
basis\textquotedblright$^{[28]}$ $e^{M}(M=0,+1,-1)$ $:$%
\begin{equation}
e^{0}=dx^{3},e^{+1}\equiv-\frac{1}{\sqrt{2}}(dx^{1}+idx^{2}),e^{-1}=\frac
{1}{\sqrt{2}}(dx^{1}-idx^{2}), \tag{A5}%
\end{equation}
$\ $ to the local comoving frames on $S^{2}:$
\begin{equation}
\mathfrak{e}^{T_{+}}=-\frac{1}{\sqrt{2}}re^{-i\gamma}\left(  d\theta
+\mathrm{i}\sin\theta d\varphi\right)  ,\mathfrak{e}^{^{T_{-}}}=\frac{1}%
{\sqrt{2}}re^{i\gamma}\left(  d\theta-\mathrm{i}\sin\theta d\varphi\right)  ,
\tag{A6}%
\end{equation}
together with the normal vector $\mathfrak{e}^{r}=dr$ and time $\mathfrak{e}%
^{t}=dt,$ $\gamma=\mp\varphi$ in the north (south) patch. In the spherical
basis $\mathfrak{e}^{m}(m=0,+1,-1),\mathfrak{e}^{0}\equiv\mathfrak{e}%
^{r},\mathfrak{e}^{\pm1}\equiv\mathfrak{e}^{^{T_{\pm}}},$ we have
\begin{equation}
\mathfrak{e}^{m}=\underset{M=0,\pm1}{\sum}D_{Mm}^{1}(\varphi,\theta
,\gamma)e^{M}, \tag{A7}%
\end{equation}
change the basis of the fixed frame $SU(2)$ generator from the cartesian basis
$T^{a}$ ($a=1,2,3$) to that in the spherical basis $T^{M}$, then gauge
transform to the comoving frame $\mathbf{T}^{m}(x)$%
\begin{equation}
\mathbf{T}^{m}(x)=\underset{M=0,\pm1}{\sum}D_{Mm}^{1}(\varphi,\theta
,\gamma)T^{M} \tag{A8}%
\end{equation}
e.g.
\[
\mathbf{T}^{r}(x)=\mathbf{T}^{0}(x)=\underset{}{\underset{a=1,2,3}{\sum}%
\hat{r}^{a}T^{a}},
\]
underline $\mathbf{e}$ $\mathbf{(T)}$ denotes the vector (generator) in
comoving frame, here the letters from the beginning ($a,b,c=1,2,3$) and from
the middle ($M,m=-1,0,+1$) of the alphabet denote the indices of orthogonal
basis and spherical basis respectively. Capital letter ($M=-1,0,+1)$ and
Lowercase ($m=-1,0,+1)$denote the indices of the fixed frame and the comoving
frame respectively. Gauge transform $A_{\mu}(x)$ from the fixed gauge to the
comoving gauge, The connection $H_{\mu}$ in (A3) gauge transform to the $U(1)$
Dirac potential (A9) while matrix tensor $K_{\mu}$ transform covariantly to
(A10)$,$%
\begin{align}
\mathbf{H}_{T_{\pm}}^{0}  &  =\mathbf{A}_{T_{\pm}}^{0}=\frac{\mp\frac
{\partial\gamma}{\partial\varphi}+\cos\theta}{r\sin\theta},\tag{A9}\\
-\mathbf{K}_{T_{+}}^{-}(x)  &  =-\mathbf{A}_{T_{+}}^{-}(x)=\mathbf{K}_{T_{-}%
}^{+}(x)=\mathbf{A}_{T_{-}}^{+}(x)=-\frac{i}{\sqrt{2}}\frac{F(r)}%
{r},\mathbf{K}_{t}^{0}(x)=\mathbf{A}_{t}^{0}(x)=G(r), \tag{A10}%
\end{align}%
\begin{equation}
A(x)=\underset{%
\begin{array}
[c]{c}%
m=0,\pm1\\
n=T_{+},T_{-},t
\end{array}
}{\sum}\mathbf{A}_{n}^{m}(x)\mathbf{e}^{n}(x)\mathbf{T}^{m}(x). \tag{A11}%
\end{equation}
Separate $F_{\mu\nu}$ into the radial and the tangential electric and magnetic
components
\begin{align}
F_{i0}^{a}  &  =E_{r}(r)x_{i}x^{a}/r^{2}+E_{T}(r)(\delta_{i}^{a}-x_{i}%
x^{a}/r^{2}),\tag{A12}\\
\varepsilon_{ijk}F_{ij}^{a}  &  =M_{r}(r)x_{k}x^{a}/r^{2}+M_{T}(r)(\delta
_{k}^{a}-x_{k}x^{a}/r^{2}).\nonumber
\end{align}
From eq. (A11) for $A(x)$, it is easy to see that the radial component
$M_{r}(x)$ can be find by using the generalize Gauss equation (C9) for the
reduced curvature;%
\begin{equation}
M_{r}(x)=2\mathbf{K}_{+}^{-}\mathbf{K}_{-}^{+}+F_{+-}^{(H)}.\mathbf{T}%
^{0}=\frac{F^{2}(r)-1}{r^{2}}, \tag{A13}%
\end{equation}
where the $F_{+-}^{(H)}=\frac{1}{r^{2}}\mathbf{T}^{0}$ is the magnetic field
of the Dirac monopole (A9) $\mathbf{T}^{a}.\mathbf{T}^{b}\equiv\frac{1}%
{2}Tr\mathbf{T}^{a}.\mathbf{T}^{b}$; the radial $E_{r}(x)$ and the tangential
$E_{T}(x),$ $M_{T}(x)$ can be find by using the generalize Codazzi equation
(C10)
\begin{equation}
E_{r}(x)=G^{\prime}(r),\qquad E_{T}(x)=\frac{G(r)F(r)}{r},\qquad
H_{T}(x)=\frac{F^{\prime}(r)}{r}. \tag{A14}%
\end{equation}
The normalizable solution of the anti-self-dual equation
\begin{equation}
F^{\prime}(r)=G(r)F(r),\qquad G^{\prime}(r)=\frac{F^{2}(r)-1}{r^{2}},
\tag{A15}%
\end{equation}
is%
\begin{equation}
F(r)=\frac{\mu r}{\mathtt{sh}\left(  \mathtt{\mu}r\right)  },\qquad
G(r)=\frac{1}{r}-\mu\mathtt{cth(\mu}r), \tag{A16}%
\end{equation}
which is holomorphic, and satisfies the condition $E_{r}=M_{r}\sim
O(1)\Rightarrow$ $F(r)\sim O(1)$, $G(r)\sim O(\frac{1}{r})$ as $r\rightarrow
0$, and asymptotically $E_{r}=M_{r}\sim\frac{1}{r^{2}}$, $E_{T}=M_{T}\sim
O\left(  \frac{1}{r^{2}}\right)  \Rightarrow G\sim1$, $F(r)\sim p(r)e^{-\mu
r},$ here $p(r)$ is a polynomial.

\section*{Appendix B. The zero mode of fermion in static spherical symmetric
BPS monopole}

Dirac equation
\begin{equation}
\gamma_{_{\mu}}D^{\mu}\Psi(x)=0, \tag{B1}%
\end{equation}
where $D_{\mu}=\partial_{\mu}-i\underset{}{\sum_{a}T^{a}A_{\mu}^{a}}$ with
$A_{\mu}$ given by (A1) and (A2); $\gamma_{j}=\left(
\begin{array}
[c]{cc}%
0 & -i\sigma_{j}\\
i\sigma_{j} & 0
\end{array}
\right)  ,\gamma_{0}=-i\left(
\begin{array}
[c]{cc}%
0 & I\\
I & 0
\end{array}
\right)  ,$

\noindent$\gamma_{5}=\left(
\begin{array}
[c]{cc}%
I & 0\\
0 & -I
\end{array}
\right)  \ $in the $\gamma_{5}$ diagonal representation.

Let $\Psi(x)=\left(
\begin{array}
[c]{c}%
\chi^{+}\\
\chi^{-}%
\end{array}
\right)  $ be static, then we get two decoupled 2-component equations with
opposite chirality.
\begin{equation}
(\sigma_{j}D_{j}\mp G(r)\mathbf{T}^{r}(x))\chi^{\pm}=0. \tag{B2}%
\end{equation}
Introduce the spin shift operator $\kappa$ by couple the spin with the orbital
momentum
\begin{equation}
\kappa=-\epsilon_{ijk}\sigma_{i}x_{j}D_{k}^{(H)}+1, \tag{B3}%
\end{equation}
then%
\begin{equation}
\sigma_{i}D_{i}^{(H)}=\sigma_{r}(\frac{\partial}{\partial r}+\frac{1}%
{r})-\frac{1}{r}\sigma_{r}\kappa,\quad\sigma_{r}\equiv\underset{i}{\sum}%
\sigma_{i}\hat{r}_{i.} \tag{B4}%
\end{equation}
So in the Dirac operator besides the spin shift operator $\kappa$ only the
radial derivative remains.%

\begin{equation}
\lbrack\sigma_{r}(\frac{\partial}{\partial r}+\frac{1}{r})-\frac{1}{r}%
\sigma_{r}\kappa-i\epsilon_{ijk}\frac{F(r)}{r}\hat{r}_{i}\sigma_{j}T_{k}\mp
G(r)\mathbf{T}^{r}]\chi^{\pm}=0. \tag{B5}%
\end{equation}
In the spherical symmetric coordinate, let%

\begin{equation}
\chi^{\pm}=\underset{J}{\sum}\underset{\mu,\nu}{\sum}f_{\mu,\nu}^{\pm
J}(r)\mathbb{D}_{\mu,\nu}^{J}(\varphi,\theta,\gamma), \tag{B6}%
\end{equation}%
\begin{equation}
\mathbb{D}_{\mu,\nu}^{J}(\varphi,\theta,\gamma)\equiv\sqrt{\frac{2J+1}{4\pi}%
}D_{M,\mu+\nu}^{J}s_{\mu}i_{\nu},\quad M=-(2J+1),\cdots,2J+1,\text{no
summation for }\mu,\nu. \tag{B7}%
\end{equation}%
\begin{equation}
s_{\mu}(\varphi,\theta,\gamma)=\underset{\mu^{\prime}}{\overset{}{\sum}}%
\sqrt{\frac{1}{2\pi}}D_{\mu^{\prime},\mu}^{S}(\varphi,\theta,\gamma
)S_{\mu^{\prime}},\quad\sigma_{3}S_{\mu}=2\mu S_{\mu},S=\frac{1}{2} \tag{B8}%
\end{equation}%
\begin{equation}
i_{\nu}(\varphi,\theta,\gamma)=\underset{\nu^{\prime}}{\overset{}{\sum}}%
\sqrt{\frac{1}{2\pi}}D_{\nu^{\prime},\nu}^{I}(\varphi,\theta,\gamma
)I_{\nu^{\prime}},\quad T_{3}I_{\nu}=2\nu I_{\nu},I=\frac{1}{2} \tag{B9}%
\end{equation}
and
\begin{equation}
\sigma_{r}s_{\mu}=2\mu s_{\mu},\qquad T_{r}i_{\nu}=\nu i_{\nu} \tag{B10}%
\end{equation}
$\sigma_{m}(x)=\underset{M=0,\pm1}{\sum}D_{Mm}^{1}(\varphi,\theta
,\gamma)\sigma^{M},\\
\mathbf{\sigma}^{r}(x)=\sigma^{0}(x)=\underset{}%
{\underset{a=1,2,3}{\sum}\hat{r}^{a}\sigma^{a}},$ or $(\sigma^{a}%
(x))_{\lambda\nu}=D_{\lambda\mu}^{\frac{1}{2}}(\varphi,\theta,\gamma
)(\sigma^{a})_{\mu\rho}D_{\rho\nu}^{\ast\frac{1}{2}}(\varphi,\theta,\gamma)$

\noindent here as in (A8) for the isospin generator $\mathbf{T}$, we
introduce the spin generators
$\sigma_{1}(x),\sigma_{2}(x),\sigma_{3}(x)$ in the comoving frames,
later when we consider the spherical case, we always write
$\sigma_{m}$ simply, by omitting the argument $x$.

Then the $J=0,1$ component $\mathbb{D}_{\mu,\nu}^{J}(\varphi,\theta,\gamma)$
of the wave function $\chi$ satisfy%

\[
J^{2}\mathbb{D}_{\mu,\nu}^{J}=J(J+1)\mathbb{D}_{\mu,\nu}^{J},\quad\quad J=0,1
\]

\[
J_{3}\mathbb{D}_{\mu,\nu}^{J}=M\mathbb{D}_{\mu,\nu}^{J},\quad
M=-(2J+1),...,2J+1
\]

\[
\sigma_{r}\mathbb{D}_{\mu,\nu}^{J}=2\mu\mathbb{D}_{\mu,\nu}^{J},\quad\mu
,\nu=\pm\frac{1}{2}%
\]

\[
T_{r}\mathbb{D}_{\mu,\nu}^{J}=\nu\mathbb{D}_{\mu,\nu}^{J},
\]

\[
\kappa\mathbb{D}_{\mu,\nu}^{J}=\kappa_{\mu}\mathbb{D}_{-\mu,\nu}^{J},
\]

\[
\varepsilon_{ijk}\hat{r}_{i}\sigma_{j}\mathbf{T}_{k}\mathbb{D}_{\mu,\nu}%
^{J}=-i2\mu\alpha_{\nu+\frac{1}{2}+\mu}^{\frac{1}{2}}\mathbb{D}_{-\mu,\nu
+2\mu}^{J},
\]
where
\begin{equation}
\alpha_{\nu}^{I}=(I+\nu)^{\frac{1}{2}}(I-\nu+1)^{\frac{1}{2}},\kappa_{\nu
}=\alpha_{\left\vert \nu\right\vert +\frac{1}{2}}^{J}=\sqrt{(J+\frac{1}%
{2})^{2}-\nu^{2}.} \tag{B11}%
\end{equation}
So the radial components $f^{\pm J}(r)$ satisfy%
\begin{equation}
(\frac{\partial}{\partial r}+\frac{1}{r})f_{\mu,\nu}^{\pm J}(r)-\frac
{\kappa_{\mu}}{r}f_{-\mu,\nu}^{\pm J}(r)+\alpha_{\nu+\frac{1}{2}+\mu}^{I}%
\frac{F(r)}{r}f_{-\mu,\nu+2\mu}^{\pm J}(r)\mp2\mu\nu G(r)f_{\mu,\nu}^{\pm
J}(r)=0. \tag{B12}%
\end{equation}
The convergence at $r\rightarrow\infty,$ requires $\mp2\mu\nu G(r)>0$
asymptotically. Thus for the

$\gamma^{5}=1\quad(-1)$ solutions $f_{\mu,\nu}^{+J}\quad(f_{\mu,\nu}^{-J})$,
$\mu\nu$ always $>0$ $(<0).$ But then the $\frac{\kappa_{\nu}}{r}f_{-\mu,\nu
}^{\pm J}$ term disobey this condition, so we should require $\kappa_{\mu}=0$
. Thus $J$ is restricted to be $0.$The superscript $J$ will be dropped later.
This implies the well-known fact that for the zero mode, the total spin, which
is contributed by the space spin and the isospin induced by the field,
cancels. Based on the same reason in section 3, we simply use the $J=0$
component in the tensor product of $S$ and $T$. From $F(r)f_{-\mu,\nu+2\mu
}^{\pm}$ term , $-\mu(\nu+2\mu)$ should be $>0$ $(<0)$ for $f^{+}(f^{-})$.
Hence $\nu=-\mu$ and $f^{+}=0$, only $f^{-}$ is convergent, it satisfies
\[
(\frac{\partial}{\partial r}+\frac{1}{r})f_{\mu,-\mu}^{-}(r)-\frac{F(r)}%
{r}f_{-\mu,\mu}^{-}(r)-\frac{1}{2}G(r)f_{\mu,-\mu}^{-}(r)=0
\]
i.e.%
\begin{align}
(\frac{\partial}{\partial r}+\frac{1}{r})f_{\frac{1}{2},-\frac{1}{2}}%
^{-}(r)-\frac{F(r)}{r}f_{-\frac{1}{2,},\frac{1}{2}}^{-}(r)-\frac{1}%
{2}G(r)f_{\frac{1}{2},-\frac{1}{2}}^{-}(r)  &  =0,\nonumber\\
(\frac{\partial}{\partial r}+\frac{1}{r})f_{-\frac{1}{2,},\frac{1}{2}}%
^{-}(r)-\frac{F(r)}{r}f_{\frac{1}{2},-\frac{1}{2}}^{-}(r)-\frac{1}%
{2}G(r)f_{-\frac{1}{2,},\frac{1}{2}}^{-}(r)  &  =0. \tag{B13}%
\end{align}
The unique convergent solution is
\begin{equation}
f^{-}(r)=f_{\frac{1}{2},-\frac{1}{2}}^{-}(r)=-f_{-\frac{1}{2},\frac{1}{2}}%
^{-}(r)\simeq r^{-\frac{1}{2}}(sh\frac{\beta r}{2})^{\frac{1}{2}}%
(ch\frac{\beta r}{2})^{-\frac{3}{2}}.\nonumber
\end{equation}
which satisfy%
\begin{equation}
(\frac{\partial}{\partial r}+\frac{1}{r})f^{-}(r)+\frac{F(r)}{r}f^{-}%
(r)-\frac{1}{2}G(r)f^{-}(r)=0. \tag{B14}%
\end{equation}
In fact, we have a local (comoving) homotopy isomorphism between the spin
$s_{\mu}$ and isospin $i_{\nu}$ in the reduced cup product $\underline
{2}\overset{\cdot}{\underset{ST}{\otimes}}\underline{2}\overset{cup~product}%
{\Longrightarrow}\underline{1}$, in the tensor product $s_{\mu}\otimes i_{\nu
}$.
\begin{equation}
\underline{2}\underset{ST}{\otimes}\underline{2}=\underline{1}\oplus
\underline{3}, \tag{B15}%
\end{equation}
the nondegenerate \textbf{zero mode }lies in $\underline{1}$ only. So we
simply have
\begin{equation}
\chi^{-}=\sum_{\mu\nu}f_{\mu\nu}^{-}s_{\mu}^{-}i_{\nu}(-1)^{\mu-\frac{1}{2}%
}\delta_{\mu,-\nu} \tag{B16}%
\end{equation}
by using%

\begin{align}
&  \frac{1}{4\pi}\underset{\mu^{\prime},\nu^{\prime}}{\sum}C_{\mu^{\prime}%
,\nu^{\prime},M}^{S,T,J}D_{\mu^{\prime}\mu}^{S}(\varphi,\theta,\gamma
)D_{\nu^{\prime}\nu}^{T}(\varphi,\theta,\gamma)=\frac{1}{\sqrt{2\pi}}%
D_{M,\mu+\nu}^{J}(\varphi,\theta,\gamma)C_{\mu,\nu,0}^{S,T,J}=(-1)^{\mu
-\frac{1}{2}}\delta_{\mu,-\nu},\tag{B17}\\
&  S=T=\frac{1}{2},J=0.\nonumber
\end{align}
Here, the finite rotation matrices $D_{\mu^{\prime}\mu}^{S}$ and
$D_{\nu^{\prime}\nu}^{T}$ from the fixed frames to the comoving frames is
coupled to a singlet expressed by the well known Kroneker matrix
$(-1)^{\mu-\frac{1}{2}}\delta_{\mu,-\nu}.$

The fixed frame Dirac eq (B5) turns to be the eq.(B12) in the \textbf{tensor
product comoving frame}, the last two terms comes from (B12)
\begin{align}
&  \gamma_{i}K_{i}=\gamma_{i}K_{i}^{a}T^{a}=\gamma_{i}\epsilon_{ija}%
\frac{F(r)}{r}\hat{r}_{j}T_{a},\nonumber\\
&  \gamma_{0}K_{0}=\gamma_{0}\delta^{ab}K_{0}^{a}T^{b}=\gamma_{0}\delta
^{ab}iG(r)\hat{r}^{a}T^{b}.\nonumber
\end{align}
And further turn to be that in the\textbf{\ cup product form}, that is
\begin{align}
&  \chi^{-}\ \text{part: }(i\epsilon_{3jk}\mathbb{\sigma}_{j}\mathbf{T}%
_{k}\frac{F(r)}{r})\chi^{-}=i(\mathbb{\sigma}_{1}\mathbf{T}_{2}-\mathbb{\sigma
}_{2}\mathbf{T}_{1})\frac{F(r)}{r}\chi^{-}\overset{cup~product}%
{\Longrightarrow}\frac{i}{2}(\mathbf{E}_{1}^{2}-\mathbf{E}_{2}^{1})\frac
{F(r)}{r}|f^{-}\rangle,\nonumber\\
&  G(r)\hat{r}^{a}T^{a}\chi^{-}\overset{cup~product}{\Longrightarrow}\frac
{1}{2}G(r)(\mathbf{E}_{1}^{1}-\mathbf{E}_{2}^{2})|f^{-}\rangle,\nonumber
\end{align}
where the cup product basis matrix $(\mathbf{E}_{i}^{j})_{\alpha\beta}%
=\delta_{i\alpha}\delta_{j\beta}$. Notice that the radial gradient term has a
$\sigma_{r}$, which turns to $\frac{1}{2}(\mathbf{E}_{1}^{1}-\mathbf{E}%
_{2}^{2})$ also, so we have at last the zero mode equation in cup product form
(B13)
\begin{align}
&  \left(  (\frac{\partial}{\partial r}+\frac{1}{r})I-\frac{F(r)}{2r}\left(
\begin{array}
[c]{cc}%
0 & 1\\
1 & 0
\end{array}
\right)  -\frac{1}{2}G(r)I\right)  |f^{-}\rangle=0,\nonumber\\
&  |f^{-}\rangle=\frac{1}{2}\left(
\begin{array}
[c]{cc}%
1 & -1\\
-1 & 1
\end{array}
\right)  \left(
\begin{array}
[c]{c}%
f_{\frac{1}{2},-\frac{1}{2}}^{-}(r)\\
f_{-\frac{1}{2},\frac{1}{2}}^{-}(r)
\end{array}
\right)  =\frac{1}{2}\left(
\begin{array}
[c]{c}%
f_{\frac{1}{2},-\frac{1}{2}}^{-}(r)-f_{-\frac{1}{2},\frac{1}{2}}^{-}(r)\\
-f_{\frac{1}{2},-\frac{1}{2}}^{-}(r)+f_{-\frac{1}{2},\frac{1}{2}}^{-}(r)
\end{array}
\right)  \tag{B18}%
\end{align}
Here we have factorize out the spherical dependence of the wave function in
the reduced cup product frame
\begin{equation}
s_{-}^{-}i_{+}\quad-\quad s_{+}^{-}i_{-},\nonumber
\end{equation}
(the superscript$-$ of $s$ denote the chirality).

By(A10) we have
\begin{equation}
\frac{i}{2}(\mathbf{E}_{1}^{2}-\mathbf{E}_{2}^{1})\frac{F(r)}{r}=\frac{1}%
{2}(\mathbf{E}_{+}^{-}-\mathbf{E}_{-}^{+})\frac{F(r)}{r}\Rightarrow\frac{1}%
{2}(\mathbf{K}_{+}^{-}-\mathbf{K}_{-}^{+})=\mathbf{K}_{+}^{-}.\nonumber
\end{equation}
so at last (B18) turns to be
\begin{equation}
((\frac{\partial}{\partial r}+\frac{1}{r})+\mathbf{K}_{T^{+}}-\frac{1}%
{2}\mathbf{K}_{t})|f^{-}\rangle=0\nonumber
\end{equation}
i.e. the eq. (B14).

Remark: Following Nahm$^{[27]}$ by Fourier transform to the momentum
space, but different from Nahm, we adopt the light cone spherical
frame coordinates, then the $|f^{-}\rangle$ becomes the holomorphic
sheaf, i.e. the mini-twistor [26 p584]. Here, the geodesic flow
along the spectral line, left real null
line, $\nabla_{r}-\frac{1}{2}\mathbf{K}_{t}\sim\nabla_{U}-i\Phi$%
[29], and the $\mathbf{K}_{T^{+}}\sim\nabla_{x}+i\nabla_{y}%
=\bar{\partial}$ [29], spans the left null plane ($\alpha$ plane),
as the natural flat connection on it.

\section*{Appendix C. The generalized Gauss Codazzi equation}

1. Condition of reducibility.

The necessary and sufficient condition for the gauge field with group $G$
defined on space-time manifold $M$ to be reducible into gauge field with
subgroup $H$, may be formulated as follows: A connection on the principal
bundle $P(M,G)$ can be reduced into the connection of subbundle $Q(M,H)$, when
and only when the associated bundle $E(M,G/H,G)$ have a section $\mathbf{n}%
(x):M\rightarrow G/H,$ which is invariant under parallel displacement. In
order to employ this condition, we must find out proper expression for $G/H$,
consequently we decompose the left invariant algebra $g$ of group $G$
canonically into $g=\mathfrak{h}+m$, where $\mathfrak{h}$ is the subalgebra
corresponding to the stationary subgroup of the element on $G/H$. Observing
the natural correspondence between $G/H$ and the subspace spanned by
$\mathfrak{h}$ in the left invariant Lie algebra, we may perform the reduction
as follows.

2. Gauge field with group $G=SU(2)$ which may be abelianized into $H=U(1)$. in
this case, $\mathfrak{h}$ is one dimensional everywhere, and its normalized
base is taken as $\hat{\mathbf{n}}$, $\hat{\mathbf{n}}\cdot\hat{\mathbf{n}%
}\equiv-2tr(\hat{\mathbf{n}}\cdot\hat{\mathbf{n}})=1$. Then the set of
$\hat{\mathbf{n}}$ makes a unit sphere $S^{2}\sim SU(2)/U(1)$ in the space of
adjoint representation. The section $\hat{\mathbf{n}}(x)$ is the mapping of
space-time $M$(except the singular point) onto $S^{2}$. This unit isospin
field is invariant under the parallel displacement by gauge potential
$\mathbf{A}_{\mu}(x)$,
\[
\tilde{\nabla}\hat{\mathbf{n}}(x)\equiv\partial_{\mu}\hat{\mathbf{n}%
}(x)+e[\mathbf{A}_{\mu}(x),\hat{\mathbf{n}}(x)]=0
\]%
\[
\mathbf{A}_{\mu}(x)\in g,\quad\mu=0,1,2,3\eqno{(C1)}
\]
where, under infinitesimal gauge transformation
\[
\mathbf{A}_{\mu}^{\prime}(x)=\mathbf{A}_{\mu}(x)+e[\mathbf{A}_{\mu}%
(x),\alpha(x)]+e\partial_{\mu}\alpha(x)
\]%
\[
\alpha(x)\in g\eqno{(C2)}
\]
Using identity $\mathbf{V}=(\mathbf{V}\cdot\mathbf{N})\mathbf{N}%
+[\mathbf{N},[\mathbf{V},\mathbf{N}]]$, we can easily see that the necessary
and sufficient condition of (C1) is
\[
\mathbf{A}_{\mu}=(\mathbf{A}_{\mu}\cdot\hat{\mathbf{n}})\hat{\mathbf{n}}%
-\frac{1}{e}[\hat{\mathbf{n}},\partial_{\mu}\hat{\mathbf{n}}]\eqno{(C3)}
\]
Here, the potential is $SU(2)$ formally, but in reality it may be transformed
at least locally into $U(1)$ potential with a constant $\hat{\mathbf{n}}(x)$
as the generator, i.e. the gauge could be chosen to turn $\hat{\mathbf{n}}(x)$
into the same direction in some region of $x$, $\partial_{\mu}\hat{\mathbf{n}%
}(x)=0$. Then $\mathbf{A}_{\mu}^{\prime}(x)$ becomes explicit Abelian, i.e.,
it equals $(\mathbf{A}^{\prime}\cdot\mathbf{n})\mathbf{n}$ in this region. If
we fix the direction of $\hat{\mathbf{n}}(x)$, but rotate a gauge angle
$\Gamma(x)$ around $\hat{\mathbf{n}}(x)$, then we obtain the $U(1)$ transform
generated by $e\hat{\mathbf{n}}:\mathbf{A}_{\mu}^{\prime}=\mathbf{A}_{\mu
}+e\hat{\mathbf{n}}\partial_{\mu}\Gamma$.

Substituting (C3) into
\[
\mathbf{F}_{\mu\nu}=\partial_{\mu}\mathbf{A}_{\nu}-\partial_{\nu}%
\mathbf{A}_{\mu}+e[\mathbf{A}_{\mu},\mathbf{A}_{\nu}]\eqno{(C4)}
\]
we get \setcounter{equation}{4}
\begin{align}
\mathbf{F}_{\mu\nu}  &  =[\partial_{\mu}(\mathbf{A}_{\nu}\cdot\hat{\mathbf{n}%
})-\partial_{\nu}(\mathbf{A}_{\mu}\cdot\hat{\mathbf{n}})]\hat{\mathbf{n}%
}-\frac{1}{e}[\partial_{\mu}\hat{\mathbf{n}},\partial_{\nu}\hat{\mathbf{n}%
}]\nonumber\\
&  =(\mathbf{F}_{\mu\nu}\cdot\hat{\mathbf{n}})\hat{\mathbf{n}} \tag{C5}%
\end{align}
In the region where $\hat{\mathbf{n}}(x)$ is well defined, $\mathbf{F}_{\mu
\nu}\cdot\hat{\mathbf{n}}$ satisfies
\[
\partial^{\mu}(^{\ast}\mathbf{F}_{\mu\nu}\cdot\hat{\mathbf{n}})=\tilde{\nabla
}^{\mu}(^{\ast}\mathbf{F}_{\mu\nu}\cdot\hat{\mathbf{n}})=(\tilde{\nabla}%
^{\mu^{\ast}}\mathbf{F}_{\mu\nu})\cdot\hat{\mathbf{n}}+^{\ast}\mathbf{F}%
_{\mu\nu}\tilde{\nabla}^{\mu}\hat{\mathbf{n}}=0\eqno{(C6)}
\]
where $\ast\mathbf{F}_{\mu\nu}\equiv\frac{1}{2}\epsilon_{\mu\nu\lambda\rho
}\mathbf{F}^{\lambda\rho}$.Locally $\mathbf{F}_{\mu\nu}\cdot\hat{\mathbf{n}}$
is the same as the ordinary electromagnetic field without magnetic charge, and
in explicit Abelian gauge it may be expressed by the $U(1)$ potential
$\mathbf{A}_{\mu}\cdot\hat{\mathbf{n}}$; $\mathbf{F}_{\mu\nu}\cdot
\hat{\mathbf{n}}=\partial_{\mu}(\mathbf{A}_{\nu}\cdot\hat{\mathbf{n}%
})-\partial_{\nu}(\mathbf{A}_{\mu}\cdot\hat{\mathbf{n}})$. But, globally its
magnetic flux through some two dimensional space like close surface
$M^{\prime}$ may be non-zero,
\[
\frac{1}{2}\iint_{M^{\prime}}\mathbf{F}_{\mu\nu}\cdot\hat{\mathbf{n}}dx^{\mu
}\wedge dx^{\nu}=\displaystyle\frac{l}{e}\iint_{S^{2}}\hat{\mathbf{n}}%
\cdot\lbrack d\hat{\mathbf{n}},\delta\hat{\mathbf{n}}]=-\frac{4\pi l}%
{e}\eqno{(C7)}
\]
Here integer $l$ is the times by which the surface $M^{\prime}$ covers the
isospin sphere $S^{2}$ through the mapping $\hat{\mathbf{n}}(x)$. Physically
it is the quantum number of the magnetic charge surrounded by the surface
$M^{\prime}$. If $l\neq0$, it is impossible to turn $\hat{\mathbf{n}}(x)$ into
one and the same direction globally by non-singular single-valued gauge
transformation. Then there must be either singularity or overlapping regions
with transition function, the corresponding Abelian potential being the
Dirac-Schwinger potential with string or the Wu-Yang global potential. Above
all, the characteristic $\pi_{1}(S^{1})$ of bundle $Q(M,H)$ of $U(1)$ gauge
field corresponds one to one to $\pi_{2}(S^{2})$ of the section $\mathbf{n}%
(x)$ on the associated coset bundle of $SU(2)$, $\pi_{1}(S^{1})\sim\pi
_{2}(S^{2})$. Their common characteristic number is determined physically by
the dual charge.Mathematically $\hat{\mathbf{n}}(x)$ is the generator of the
holonomy group of $P(M,G)$ under given connection;physically e$\hat
{\mathbf{n}}(x)$ is the charge operator;abelianizable $\mathbf{A}_{\mu}$ is
the potential of pure electromagnetic field $\mathbf{F}_{\mu\nu}$; and
meantime, $\hat{\mathbf{n}}$ is the common isodirection of the six space-time
components of $\mathbf{F}_{\mu\nu}$.

3.Non-abelianizable $SU(2)$ potential $\mathbf{A}_{\mu}(x)$ and field strength
$\mathbf{F}_{\mu\nu}(x)$. Now, the holonomy group are whole $SU(2)$, thus it
is impossible to choose from its generators some $\hat{\mathbf{n}}(x)$ which
remains invariant under parallel displacement. The noninvariant section
$\hat{\mathbf{n}}(x)$ must be given otherwise. Physically,as the charge
operator, $\hat{\mathbf{n}}(x)$ is the phase axis of wave functions of charged
particles, or the isodirection of its current vector,or $\hat{\mathbf{n}%
}(x)=\phi(x)/|\phi(x)|$, where $\phi(x)$ is the Higgs particle. In pure gauge
field without other particles, $\hat{\mathbf{n}}(x)$ may be the privileged
direction determined by the intrinsic symmetry of the field, e.g. the
generator of the stationary subgroup $H$ for $G$ invariant connection. (In
case of synchronous of space spin and isospin spherical symmetry field as the
BPS monopole, the privileged direction is synchronous with the vector radius
$\hat{\mathbf{r}}$.)

Once a section $\hat{\mathbf{n}}(x)$ is given, it determines a corresponding
subbundle $Q(M,H)$, whose characteristic class is decided by the homotopic
property of $\hat{\mathbf{n}}(x). $Physically, as soon as the charge operator
\ $\hat{\mathbf{n}}(x)$\ is given, one can separate the $U(1)$ Dirac
component\ $\mathbf{H}_{\mu}(x)$\ from the $SU(2)$ potential\ $\mathbf{A}%
_{\mu}(x)$\ as follows: Here\ $\mathbf{H}_{\mu}(x)$\ is the $U(1)$ gauge
potential with\ $\hat{\mathbf{n}}(x)$\ as the generator. Now from\ $\widetilde
{\nabla}_{\mu}\hat{\mathbf{n}}\equiv\partial_{\mu}\mathbf{n}+e[\mathbf{A}%
_{\mu},\hat{\mathbf{n}}]$ \ which is not vanishing now, we get
\[
\mathbf{A}_{\mu}=\displaystyle(\mathbf{A}_{\mu}\cdot\hat{\mathbf{n}}%
)\hat{\mathbf{n}} -\frac{1}{e}[\hat{\mathbf{n}},\partial_{\mu}\hat{\mathbf{n}%
}]+\frac{1}{e}[\hat{\mathbf{n}},\nabla_{\mu}\hat{\mathbf{n}}] \equiv
\mathbf{H}_{\mu}+\mathbf{K}_{\mu}\eqno{(C8)}
\]
Here we have set$\displaystyle \mathbf{H}_{\mu}\equiv(\mathbf{A}_{\mu}%
\cdot\hat{\mathbf{n}})\hat{\mathbf{n} }-\frac{1}{e}[\hat{\mathbf{n}}%
,\partial_{\mu}\hat{\mathbf{n}}]$.It is easy to prove that $\mathbf{H}_{\mu}$
satisfies(C1)-(C3),hence it is the U(1) part in $\mathbf{A}_{\mu}$, the
remainder $\frac{1}{e}[\hat{\mathbf{n}},\nabla_{\mu}\hat{\mathbf{n}}%
]\equiv\mathbf{K}_{\mu}$ is gauge covariant and represents the charged vector
particles. Geometrically $e\mathbf{K}_{\mu}$corresponds to the second
fundamental form, e.g.$\nabla_{\mu}\hat{\mathbf{n}}=[e\mathbf{K}_{\mu}%
,\hat{\mathbf{n}}]$ is the generalized Weingarten formula (since
$|\mathbf{\hat{\mathbf{n}}}|=1$, the "normal" component is absent).
Substituting (C8) into $\mathbf{F}_{\mu\nu}=\partial_{\mu}\mathbf{A}_{\nu
}-\partial_{\nu}\mathbf{A}_{\mu}+e[\mathbf{A}_{\mu},\mathbf{A}_{\nu}]$ and
making comparison with (C5). We get
\[%
\begin{array}
[c]{rcl}%
\mathbf{F}_{\mu\nu}\cdot\hat{\mathbf{n}}\ \hat{\mathbf{n}}-[\mathbf{K}_{\mu
},\mathbf{K}_{\nu}] & = & F_{\mu\nu}\cdot\hat{\mathbf{n}}\ \hat{\mathbf{n}%
}-[\nabla_{\mu}\hat{\mathbf{n}},\nabla_{\nu}\hat{\mathbf{n}}]\\
& = & \displaystyle\partial_{\mu}(\mathbf{H}_{\nu}\cdot\hat{\mathbf{n}}%
)\hat{\mathbf{n}} -\partial_{\nu}(\mathbf{H}_{\mu}\cdot\hat{\mathbf{n}}%
)\hat{\mathbf{n}}-\frac{1}{e}[\partial_{\mu}\hat{\mathbf{n}} ,\partial_{\nu
}\hat{\mathbf{n}}]\\
& = & \mathbf{F}^{(H)}_{\mu\nu}\hspace*{260pt}(C9)
\end{array}
\]
This is just the 't Hooft expression. Here $\mathbf{F}^{(H)}_{\mu\nu}$ is the
$U(1)$ field part in $\mathbf{F}_{\mu\nu}$contributed by $\mathbf{H}_{\mu}$
(The Higgs particle does not contribute this $U(1)$ field, only its
isodirection coincides with that of charge operator.) Geometrically (C9) is
the generalized Gauss equation. $\mathbf{F}^{(H)}_{\mu\nu}\cdot\hat
{\mathbf{n}}$, as the $U(1)$ "subcurvature" of the total curvature
$\mathbf{F}_{\mu\nu}$, satisfies the Bianchi identity on subbundle. At the
same time we get the generalized Codazzi equation,
\[
[\hat{\mathbf{n}},[\mathbf{F}_{\mu\nu},\hat{\mathbf{n}}]] =\widetilde{\nabla
}_{\mu}\mathbf{K}_{\nu}-\widetilde{\nabla}_{\nu}\mathbf{K}_{\mu}\eqno{(C10)}
\]

\end{document}